\documentstyle[osa,twocolumn]{revtex}

\begin{document}
\title{Scaling picture of magnetism formation in the anomalous f-systems: interplay
of the Kondo effect and spin dynamics}
\author{V.Yu.Irkhin and M.I.Katsnelson$^*$}
\address{Institute of Metal Physics, 620219 Ekaterinburg, Russia}
\maketitle

\begin{abstract}
Formation of magnetically ordered state in the Kondo lattices is treated
within the degenerate $s-f$ exchange and Coqblin-Schrieffer models. The
Kondo renormalizations of the effective coupling parameter, magnetic moment
and spin excitation frequencies are calculated within perturbation theory.
The results of one-loop scaling consideration of the magnetic state in Kondo
lattices are analyzed. The dependence of the critical values of the bare
model parameters on the type of the magnetic phase and space dimensionality
is investigated. Renormalization of the effective Kondo temperature by the
interatomic exchange interactions is calculated. An important role of the
character of spin dynamics (existence of well-defined magnon excitations,
presence of magnetic anisotropy etc.) is demonstrated. The regime of
strongly suppressed magnetic moments, which corresponds to magnetic
heavy-fermion system, may occur in a rather narrow parameter region only. At
the same time, in the magnetically ordered phases the renormalized Kondo
temperature depends weakly on the bare coupling parameter in some interval.
The critical behavior, corresponding to the magnetic transition with
changing the bare $s-f$ coupling parameter, is investigated. In the vicinity
of the strong coupling regime, the spectrum of the Bose excitations becomes
softened. Thus on the borderline of magnetic instability the Fermi-liquid
picture is violated in some temerature interval due to scattering of
electrons by these bosons. This may explain the fact that a non-Fermi-liquid
behavior often takes place in the heavy-fermion systems near the onset of
magnetic ordering.
\end{abstract}


\section{Introduction}

Anomalous 4$f$- and 5$f$-compounds, including so-called Kondo lattices and
heavy-fermion systems, are studied extensively starting from the middle of
80s \cite{Stewart,Brandt,Adr}. From very beginning of these investigations
it became clear that the effects, connected with regular arrangement of the
Kondo centres (rare-earth or actinide ions) play a crucial role in the
physics of such systems. When passing from one magnetic centre to the Kondo
lattice, two main new features appear. First, provided that the strong
coupling regime takes place, the Abrikosov-Suhl resonance in the one-site $t$%
-matrix leads to formation of a complicated band structure near $E_F$ on the
new energy scale (the Kondo temperature $T_K)$ with sharp peaks and
pseudogaps in the density of states\cite{Fulde,Zw}. This provides a common
explanation of the heavy-fermion behavior. Second, the competition between
the Kondo screening of magnetic moments and intersite magnetic interactions
has a great importance \cite{Lee,Zw,Fulde,Lac}. Following to the old paper
by Doniach \cite{Don}, it was believed in early works that this competition
leads to total suppression of either magnetic moments or the Kondo
anomalies. However, more recent experimental data and careful theoretical
investigations made clear that the Kondo lattices {\it as a rule }%
demonstrate magnetic ordering or are close to this. This concept was
consistently formulated and justified in a series of our papers \cite
{IKFTT,IKZ1,Von1,Von2,IKJP92}. A very important circumstance is that
interspin coupling between the Kondo sites results in a smearing of
singularities in electron and magnetic properties on the scale of the
characteristic spin-dynamics frequency $\overline{\omega }.$ At the same
time, $\overline{\omega }$ itself acquires renormalizations resulting in its
decrease due to the Kondo screening. A simple scaling consideration of this
renormalization process in the $s-f$ exchange model \cite{IKJP92} yields,
depending on the values of bare parameters, both the ``usual'' states (a
non-magnetic Kondo lattice or a magnet with weak Kondo contributions) and
the peculiar magnetic Kondo-lattice state. In the latter state, small
variations of parameters result in strong changes of the ground-state
moment. Thus a characteristic feature of heavy fermion magnets --- high
sensitivity of the ground-state moment to external factors like pressure and
doping by a small amount of impurities --- is naturally explained. At the
same time, only the simplest $s-f$ model was considered in Ref.\cite{IKJP92}%
, and the equations obtained were not investigated in detail. Therefore a
number of important features of the Kondo magnets were not described.

Recently, a number of anomalous $f$-systems (U$_x$Y$_{1-x}$Pd$_3$, UPt$%
_{3-x} $Pd$_x$, UCu$_{5-x}$Pd$_x$, CeCu$_{6-x}$Au$_x$, U$_x$Th$_{1-x}$Be$%
_{13}$ etc.) demonstrating the so-called non-Fermi-liquid (NFL) behavior
have become a subject of great interest (see, e.g., the reviews \cite{Maple}%
). It should be noted that such a behavior is observed not only in alloys,
but also in some stoichiometric compounds, e.g., Ce$_7$Ni$_3$ \cite{CeNi},
CeCu$_2$Si$_2,$CeNi$_2$Ge$_2$ \cite{Steg}. These systems possess unusual
logarithmic or power-law temperature dependences of electron and magnetic
properties. It is a common practice to discuss such a behavior within the
one-impurity two-channel Kondo model \cite{Tsv,Col}. However, the NFL
behavior is typical for systems lying on the boundary of magnetic ordering
and demonstrating strong spin fluctuations \cite{Maple}. So, many-center
effects should play an important role in this phenomenon. At the same time,
for a number anomalous $f$-systems as Sm$_3$Se$_4,$ Yb$_4$As$_3,$ as well as
for the only ``moderately heavy-fermion'' $d$-system Y$_{1-x}$Sc$_x$Mn$_2,$
the heavy-fermion state itself seems to be connected with peculiarities of
intersite couplings (e.g., frustrations), rather than with the one-impurity
Kondo effect\cite{PL,Von1,Von2}. Thus the interplay of the Kondo effect and
intersite spin dynamics results in very rich and complicated picture rather
than in trivial mutual suppression.

The aim of the present paper is a systematic investigation of formation of
the magnetic Kondo-lattice state and of its properties for various magnetic
phases depending on the character of spin dynamics.

In Sect.2 we introduce main theoretical models which enable one to treat the
Kondo effect in a lattice in different cases with account of orbital
degeneracy. In Sect.3 we analyze the properties of the localized spin
subsystem for these models in the absence of $s-f$ interaction. The Kondo
renormalizations in the paramagnetic state with account of spin dynamics are
considered in Sect.4. In Sects.5 and 6 we calculate the Kondo corrections to
the electron spectrum (and thereby to the effective $s-f$ coupling) and
magnon frequencies in the magnetically ordered phases. In Sect.7 we write
down the scaling equations for the effective $s-f$ parameter and
spin-fluctuation frequency which plays the role of a cutoff for the Kondo
divergences in concentrated $f$-systems. In Sect.8 we generalize the scaling
approach for the ordered state by including the renormalization of the
residue of the spin Green's function at the magnon pole. The most simple
large-$N$ limit in the Coqblin-Schrieffer model where spin dynamics is
unrenormalized is considered in Sect.9. The scaling picture for finite $N$
is discussed in Sect.10. In Sect.11 we discuss the critical behavior near
the magnetic phase transition and discuss a possibility of the Fermi-liquid
picture violation. The scaling behavior in the $s-f$ exchange model with a
large orbital degeneracy is investigated in Sect.12, and an explicit
description of the non-Fermi-liquid behavior is obtained in this limiting
case.

\section{ Theoretical models}

To treat the Kondo effect in a lattice we use the $s-d(f)$ exchange
Hamiltonian 
\begin{equation}
H=\sum_{{\bf k}\sigma }t_{{\bf k}}c_{{\bf k}\sigma }^{\dagger }c_{{\bf k}%
\sigma }^{}+H_f+H_{sf}=H_0+H_{sf}  \label{H}
\end{equation}
where $t_{{\bf k}}$ is the band energy. We consider the pure spin $s-d(f)\,$%
exchange model with 
\begin{equation}
H_f=\sum_{{\bf q}}J_{{\bf q}}{\bf S}_{{\bf -q}}{\bf S}_{{\bf q}%
},\,\,H_{sf}=-\sum_{{\bf kk}^{\prime }\alpha \beta }I_{{\bf kk}^{\prime }}%
{\bf S}_{{\bf k-k}^{\prime }}\mbox {\boldmath $\sigma $}_{\alpha \beta }c_{%
{\bf k}\alpha }^{\dagger }c_{{\bf k}^{\prime }\beta }^{}  \label{sfe}
\end{equation}
where ${\bf S}_i$ and ${\bf S}_{{\bf q}}$ are spin operators and their
Fourier transforms, ${\bf \sigma }$ are the Pauli matrices. For the sake of
convenient constructing perturbation theory, we explicitly include the
Heisenberg exchange interaction with the parameters $J_{{\bf q}}$ in the
Hamiltonian, although in fact this interaction can be the indirect RKKY
coupling. Expanding the $s-d(f)$ interaction in spherical functions we have 
\begin{equation}
I_{{\bf kk}^{\prime }}=\sum_{lm}I_lY_{lm}^{*}(\theta _{{\bf k}},\phi _{{\bf k%
}})Y_{lm}(\theta _{{\bf k}^{\prime }},\phi _{{\bf k}^{\prime }})  \label{YY}
\end{equation}
Hereafter we retain in (\ref{YY}) only one term $I_l\equiv I$ ($l=2$ for $d$%
-electrons and $l=3$ for $f$-electrons). Introducing the operators 
\begin{equation}
c_{{\bf k}m\sigma }=i^l(4\pi )^{1/2}c_{{\bf k}\sigma }Y_{lm}(\theta _{{\bf k}%
},\phi _{{\bf k}})
\end{equation}
which satisfy, after averaging over the angles of the vector {\bf k}, the
Fermi commutation relations, we reduce $H_{sf}$ to the form 
\begin{equation}
H_{sf}=-I\sum_{{\bf kk}^{\prime }m\alpha \beta }{\bf S}_{{\bf k-k}^{\prime }}%
\mbox {\boldmath $\sigma $}_{\alpha \beta }c_{{\bf k}m\alpha }^{\dagger }c_{%
{\bf k}^{\prime }m\beta }^{}
\end{equation}
Assuming the electron and spin excitation spectrum to be isotropic, in final
expressions for self-energies we can perform averaging over the angles of
all the wavevectors and use the orthogonality relation 
\[
\int \int \sin \theta d\theta d\phi dY_{lm}^{*}(\theta ,\phi )Y_{lm^{\prime
}}(\theta ,\phi )=\delta _{mm^{\prime }} 
\]
Thus the factors of $[l]=(2l+1)$ occur in any order of perturbation theory,
and we have to replace $I^n\rightarrow [l]I_l^n$ in the ``connected'' terms
of perturbation expansion in comparison with the ``standard'' $s-d$ model ($%
l=0)$.

It is worthwhile to remember main results for the one-impurity version of
this model \cite{Noz,Wieg}. Perturbation theory treatment leads to
occurrence of infrared divergences. Provided that $I<0,$ the characteristic
energy scale (the Kondo temperature) 
\begin{equation}
T_K=D\exp (1/2I\rho )  \label{TK}
\end{equation}
occurs, where $D$ is of order of bandwidth, $\rho $ is the bare density of
electron states at the Fermi level with one spin projection. At $T\sim T_K$
the effective $s-f$ interaction becomes very large and the system enters the
strong coupling regime. The electron energy spectrum in this region is
determined by the presence of the Abrikosov-Suhl resonance of the width $%
T_K. $ Properties of the ground state and character of the low-temperature
behavior depend crucially on the parameters $S$ and $[l].$ At $2S=[l]$ the
Fermi-liquid singlet state occurs. At $2S>[l]$ the localized moment and
logarithmic behavior of electronic characterisitics retains, but the
replacement $S\rightarrow S-[l]/2$ takes place. At $2S<[l]$ a very
interesting ``overcompensated'' regime occurs. Last time, the particular
case of this regime with $S=1/2,\,[l]=2$ (the two-channel Kondo model
describing the non-Fermi-liquid behavior, see, e.g.,\cite{Tsv,Col}) is a
subject of great interest.

In the case of the periodic model presence or absence of the strong coupling
regime depends also on the character of intersite spin-spin interactions
which are described by $H_f.$ This factor will be analyzed in detail below.

The $s-d(f)$ model does not take into account scattering by orbital degrees
of freedom \cite{Schr,Hirst,Noz,II}. Another important model, which is used
frequently to describe the Kondo effect, is the Coqblin-Schrieffer model.
For its periodic version with the $\,f-f$ exchange interaction of the $SU(N)$
form we have 
\begin{eqnarray}
H_f &=&\frac 12\sum_{{\bf q}}J_{{\bf q}}\sum_{M,M^{\prime }=-S}^SX_{-{\bf q}%
}^{MM^{\prime }}X_{{\bf q}}^{M^{\prime }M},\,  \label{CS} \\
\,H_{sf} &=&-I\sum_{{\bf kk}^{\prime }MM^{\prime }}X_{{\bf k}^{\prime }{\bf %
-k}}^{MM^{\prime }}c_{{\bf k}^{\prime }M^{\prime }}^{\dagger }c_{{\bf k}M}^{}
\nonumber
\end{eqnarray}
where $X_{{\bf q}}^{MM^{\prime }}$ are the Fourier transforms of the
Hubbard's operators for the localized spin system 
\[
X_i^{MM^{\prime }}=|iM\rangle \langle iM^{\prime }| 
\]
$S=(N-1)/2$ is the total angular momentum (this notation is used for the
sake of convenience and has a somewhat different meaning in comparison with
the spin $S$ in the $s-f$ model), the operators 
\begin{equation}
c_{{\bf k}M}=\sum_{m\sigma }C_{\frac 12\sigma ,lm}^{SM}c_{{\bf k}m\sigma }
\end{equation}
possess, after averaging over the angles, the Fermi properties due to the
orthogonality relations for the Clebsh-Gordan coefficients $C$. As well as
for the model (\ref{sfe}), we will assume that this averaging should be
performed when calculating the Green's functions. The Hamiltonian $H_{sf}$
with $I<0$ can be derived from the degenerate Anderson-lattice model for
rare-earth compounds ($LS$-coupling); a Hamiltonian of the same form occurs
in the case of $jj$-coupling (actinide systems) \cite{II}.

For $S=1/2,N=2,\,l=0$ the models (\ref{sfe}) and (\ref{CS}) reduce to the
standard $s-f$ model with $S=1/2$ and coincide. The ground state in the
one-impurity Coqblin-Schrieffer model is similar to that in the $s-f$ model
with $2S=[l],$ i.e. complete screening of the localized moment and a
Fermi-liquid picture take place. Due to another structure of perturbation
theory for the Coqblin-Schrieffer model, we have to replace $2\rightarrow N$
in (\ref{TK}). Thus the role of the degeneracy factors in both the models
under consideration is different: the expression for $T_K$ does not contain
the factor of $[l]=(2l+1)$ in the model (\ref{sfe}), but contains the factor
of $N$ in the model (\ref{CS}). Peculiarities of the Coqblin-Schrieffer
model are determined by that the transitions between any values of localized 
$f$-state projection $M$ are possible, so that the number of excitations
branches is large. We shall see that this may result in essential
modifications of magnetic behavior.

The interaction $H_f$ in (\ref{CS}) can be obtained as an indirect RKKY-type
interaction which arises to second order in $H_{sf}$. Using the standard
Heisenberg interaction (as in (\ref{sfe})), where only $M\rightarrow M\pm 1$
transitions are allowed, is incovenient since other transitions acquire an
energy gap. However, inclusion of this interaction does not lead to a strong
change of the physical picture.\thinspace The standard angular momentum
operators on a site are expressed in terms of the $X$-operators as 
\begin{eqnarray}
S_i^{+} &=&\sum_M(S-M)^{1/2}(S+M+1)^{1/2}X_i^{M+1,M}  \label{SX} \\
S_i^z &=&\sum_MMX_i^{MM}  \nonumber
\end{eqnarray}

Of course, both the $s-f$ exchange model and Coqblin-Schrieffer model are
some idealizations of realistic situation. Choice of adequate model for a
given compound depends mainly on the relation between the width of $f$-level
and the spin-orbital coupling parameter. If the broadening of $f$-level due
to either the hybridization or direct $f-f$ overlap is larger than its
spin-orbital splitting we should consider the latter after the transition
from atom-like $f$-states to the crystal states. Usually the orbital moment
is quenched in Bloch-like states \cite{Orb} and therefore only the spin
moment should be taken into account when considering the interaction with
conduction electrons. The Coqblin-Schrieffer model was initially proposed to
describe cerium and ytterbium systems, especially diluted ones \cite{CS}.
However, as it is clear now, the situation for cerium compounds is more
complicated. First, the spin-orbit coupling for cerium is in fact not too
large (about 0.25 eV) and comparable with the width of the virtual $f$%
-level. Second, a number of cerium system, including the pure $\alpha -$%
cerium have rather large $f-f$ overlap (the relative role of $f-f$ overlap
and hybridization is discussed in detail in the review \cite{VKT}). Thus the
applicability of one of these models should be considered separately for any
specific compound. We shall demonstrate below that results of the scaling
consideration for the Coqblin-Schrieffer model and $s-f$ exchange model are
essentially different.

The model (\ref{CS}) may be generalized to include two magnetic $f$%
-configurations with the angular moments $J$ and $J^{\prime }$ so that 
\begin{equation}
H_{sf}=-I\sum C_{J^{\prime }\mu ,jm}^{JM}C_{J^{\prime }\mu ,jm^{\prime
}}^{JM^{\prime }}X_{{\bf k}^{\prime }{\bf -k}}^{MM^{\prime }}c_{{\bf k}%
^{\prime }m^{\prime }}^{\dagger }c_{{\bf k}m}^{}  \label{Ht}
\end{equation}
(we restrict ourselves for simplicity to the case of $jj$-coupling bearing
in mind uranium compounds). However, this Hamiltonian has a complicated
tensor structure \cite{Noz,II} and does not enable one to calculate an
unique energy scale by using perturbation expansion. Such an energy scale
can be obtained starting from the low-temperature regime and reads\cite{II} 
\begin{equation}
T_K=D\exp \left[ -1\left/ I\rho \left( \frac{2J+1}{2J^{\prime }+1}-1\right)
\right. \right]  \label{twom}
\end{equation}
(note that the exponent in (\ref{twom}) for the case $J^{\prime }=0$ differs
by an unity from the correct result; such a difference is typical for the
methods which are in fact based on the large-$N$ expansion \cite{New,Bick}).
In the case $J>J^{\prime }$ the situation for the model (\ref{twom}) is
similar to that for the $s-f$ model with $2S>[l].$

\section{Ground state and spectrum of spin excitations}

In the ground ferromagnetic (FM) state to zeroth approximation in $I$ we
have $\langle S_i^z\rangle =S,$ and the spin-wave spectrum for the model (%
\ref{sfe}) reads 
\begin{equation}
\omega _{{\bf q}}=\omega _{{\bf q}}^{FM}(S)=2S(J_{{\bf q}}-J_0)
\end{equation}
For the model (\ref{CS}), we also assume the magnon character of spin
dynamics. Then the spin-wave spectrum can be found by linearizing the
equation of motion for the ``spin deviation'' operator $X_{-{\bf q}}^{MS}$.
There exist $2S=N-1$ spin-wave modes with the frequency $\omega _{{\bf q}%
}=\omega _{{\bf q}}^{FM}(1/2)$, which correspond to the transitions $%
S\rightarrow M<S.$

Now we discuss an antiferromagnet which has the spiral structure along the $%
x $-axis with the wavevector {\bf Q }($J_{{\bf Q}}=J_{\min }$) 
\[
\langle S_i^z\rangle =S\cos {\bf QR}_i,\,\langle \,S_i^y\rangle =S\sin {\bf %
QR}_i,\,\langle S_i^x\rangle =0 
\]
We introduce the local coordinate system 
\begin{eqnarray}
S_i^z &=&\hat S_i^z\cos {\bf QR}_i-\hat S_i^y\sin {\bf QR}_i, \\
\,S_i^y &=&\hat S_i^y\cos {\bf QR}_i+\hat S_i^z\sin {\bf QR}_i,\,S_i^x=\hat S%
_i^x  \nonumber
\end{eqnarray}
Hereafter we consider for simplicity a two-sublattice AFM (2{\bf Q }is equal
to a reciprocal lattice vector, so that $\cos ^2{\bf QR}_i=1,\,\sin ^2{\bf QR%
}_i=0$). Passing to the spin deviation operator we derive in the spin-wave
region 
\begin{eqnarray}
H_f &=&\text{const }+\sum_{{\bf q}}[C_{{\bf q}}b_{{\bf q}}^{\dagger }b_{{\bf %
q}}+\frac 12D_{{\bf q}}(b_{-{\bf q}}b_{{\bf q}}+b_{{\bf q}}^{\dagger }b_{-%
{\bf q}}^{\dagger })]  \label{Hsw} \\
C_{{\bf q}} &=&S(J_{{\bf Q+q}}+J_{{\bf q}}-2J_{{\bf Q}}),\,D_{{\bf q}}=S(J_{%
{\bf q}}-J_{{\bf Q+q}})  \nonumber
\end{eqnarray}
Diagonalizing (\ref{Hsw}) we obtain the spin-wave spectrum 
\begin{equation}
\omega _{{\bf q}}=\omega _{{\bf q}}^{AFM}(S)=(C_{{\bf q}}^2-D_{{\bf q}%
}^2)^{1/2}=2S(J_{{\bf q}}-J_{{\bf Q}})^{1/2}(J_{{\bf Q+q}}-J_{{\bf Q}})^{1/2}
\label{waf}
\end{equation}
The corresponding transformation to the local coordinate system for the
model (\ref{CS}) has the form 
\[
X_i^{MM^{\prime }}=\frac 12[\widehat{X}_i^{M,M^{\prime }}(1+\cos {\bf QR}_i)+%
\widehat{X}_i^{-M,-M^{\prime }}(1-\cos {\bf QR}_i)] 
\]
Then we obtain 
\begin{equation}
H_f=\sum_{{\bf q}}\sum_{M,M^{\prime }=-S}^S(J_{{\bf q}}^{(2)}\widehat{X}_{-%
{\bf q}}^{MM^{\prime }}\widehat{X}_{{\bf q}}^{M^{\prime }M}+J_{{\bf q}}^{(1)}%
\widehat{X}_{-{\bf q}}^{MM^{\prime }}\widehat{X}_{{\bf q}}^{-M^{\prime }-M})
\end{equation}
\begin{eqnarray}
H_{sf} &=&-I\sum_{{\bf kq}}\sum_{M,M^{\prime }=-S}^S[(\widehat{X}_{{\bf q}%
}^{M,M^{\prime }}+\widehat{X}_{{\bf q}}^{-M,-M^{\prime }})  \nonumber \\
&&+(\widehat{X}_{{\bf q+Q}}^{M,M^{\prime }}-\widehat{X}_{{\bf q+Q}%
}^{-M,-M^{\prime }})]c_{{\bf k}M^{\prime }}^{\dagger }c_{{\bf k-q}M}^{}
\end{eqnarray}
with 
\[
J_{{\bf q}}^{(1,2)}=\frac 12(J_{{\bf q}}\mp J_{{\bf q+Q}}) 
\]
In the mean field approximation we have 
\begin{equation}
\langle H_f\rangle =\sum_{M=-S}^S(J_0^{(2)}n_M^2+J_0^{(1)}n_Mn_{-M})
\label{Hf}
\end{equation}
with $n_M=\langle \widehat{X}_i^{MM}\rangle .$ The usual AFM state turns out
to be unstable for simple lattices provided that only the nearest-neighbor
interaction is taken into account ($J_{{\bf q}}^{(2)}\equiv 0$). Indeed, in
this case any state with 
\[
\sum_{M>0}n_M=1,\sum_{M<0}n_M=0 
\]
has the same energy $\langle H_f\rangle =0.$ When introducing the
next-neighbor ferromagnetic interaction ($J_0^{(2)}<0$) the AFM state with 
\begin{equation}
n_S=1,\,n_{M<S}=0  \label{AF1}
\end{equation}
is stabilized. This state will be referred to as AFM1. The corresponding
``spin-wave'' spectrum contains $N$ branches. The mode, which corresponds to
the transition $S\rightarrow -S,$ has the frequency $\omega _{{\bf q}%
}^{(a)}=\omega _{{\bf q}}^{AFM}(S=1/2).$ Other $N-2$ modes have a
ferromagnetic type and possess the energy 
\[
\omega _{{\bf q}}^{(f)}=\omega _{{\bf q}}^{FM}(S=1/2,\,J_{{\bf q}%
}\rightarrow J_{{\bf q}}^{(2)}) 
\]

Provided that next-neighbor interaction is antiferromagnetic too ($%
J_0^{(2)}>0,$ this case is referred to as AFM2), minimization of (\ref{Hf})
yields 
\begin{equation}
n_{M<0}=0,\,n_{M>0}=K\equiv 
{2/N=2/(2S+1),\,N\text{ even} \atopwithdelims\{. J_0/(SJ_0+J_0^{(2)}),\,N\text{ odd}}
\label{AF2}
\end{equation}
For odd $N$ we have 
\begin{equation}
n_0=J_0^{(2)}/(SJ_0+J_0^{(2)})\simeq J_0^{(2)}/(SJ_0)\ll K
\end{equation}
Then, according to (\ref{SX}), for $N>2$ the sublattice magnetization turns
out to be reduced in comparison with $S$ already in the mean-field
approximation: 
\begin{equation}
\overline{S}=\frac 12\times 
{S+1/2,\,S\,\text{ half-integer} \atopwithdelims\{. (S+1)(1-n_0),\,S\text{ integer}}
\label{S0}
\end{equation}
Then there exist $N^2/4$ ($N$ even) or $(N-1)^2/4$ ($N$ odd)
``antiferromagnetic'' modes which correspond to the transitions from $M>0$
to $M^{\prime }<0$ and have the frequency $\omega _{{\bf q}}^{(a)}=\omega _{%
{\bf q}}^{AFM}(S=K)$. Besides that, for odd $N$ there exist peculiar modes
with the frequencies $\omega _{{\bf q}}^\Omega ,$ which describe the
transitions $M$ $\rightarrow $ $0$ and are determined by the equations 
\begin{equation}
\Omega _{{\bf q}}^{\pm }(\omega )\equiv (\omega \mp (K-2n_0)C_{{\bf q}%
})(\omega \pm n_0C_{{\bf q}})+n_0(K-2n_0)D_{{\bf q}}^2=0  \label{Om}
\end{equation}
(the plus sign corresponds to $M>0$ and the minus sign to $M<0),$ so that to
lowest order in $n_0$ 
\begin{equation}
\omega _{{\bf q}1}^{\Omega \pm }\simeq \pm KC_{{\bf q}},\omega _{{\bf q}%
1}^{\Omega \pm }\simeq \mp n_0\,(C_{{\bf q}}^2-D_{{\bf q}}^2)/C_{{\bf q}}
\label{om}
\end{equation}

Thus one of the solutions to (\ref{Om}) describes the mode which is very
soft for $J_0^{(2)}\ll SJ_0.$ Since the Kondo singular contributions (see
below), which are cut at this mode, are large, this may result in a tendency
to the destruction of magnetism.

Really, the considered ground states and excitation spectra are strongly
influenced by the crystal field \cite{Noz,FLow,Cox}. In particular, large
gaps and degeneracy lift may occur in the above-discussed ``additional''
modes. This question needs a special consideration.

\section{Renormalization in the paramagnetic phase}

The Kondo-lattice problem in the paramagnetic state describes the process of
screening of localized magnetic moments.

To find the renormalization of the effective $s-f$ exchange parameter we
consider the electron self-energy. In the second order in $I$ we have 
\begin{equation}
\Sigma _{{\bf k}}^{(2)}(E)=I^2P\sum_{{\bf q}}\frac 1{E-t_{{\bf k-q}}}
\label{sig2}
\end{equation}
where 
\begin{equation}
P= 
{[l]S(S+1),\text{ model (\ref{sfe})} \atopwithdelims\{. 1-1/N^2,\text{ model (\ref{CS})}}
\end{equation}
To construct a self-consistent theory of Kondo lattices we have to calculate
the third-order Kondo correction to the self energy with account of spin
dynamics. Such calculations were performed in Refs.\cite{IKFTT} within the
simplest $s-f$ model. In our case we obtain for the singular contribution
(which contains logarithmic divergence in the absence of spin dynamics) 
\begin{equation}
\Sigma _{{\bf k}}^{(3)}(E)=-I^3PN\int_{-\infty }^\infty d\omega \sum_{{\bf %
q,p}}{\cal J}_{{\bf q}}(\omega )\frac{n_{{\bf k-q}}}{E-t_{{\bf k-q}}-\omega }%
\left( \frac 1{E-t_{{\bf k-p}}}-\frac 1{t_{{\bf k-q}}-t_{{\bf k-p}}}\right)
\label{sig3}
\end{equation}
where $N=2$ in the model (\ref{sfe}),$\;{\cal J}_{{\bf q}}(\omega )$ is the
spectral density of the spin Green's function for the Hamiltonian $H_f,$
which is normalized to unity, the spin dynamics is neglected in the
denominator, which is not connected with the Fermi function $n_{{\bf k}%
}=n(t_{{\bf k}})$. The Kondo renormalization of the $s-f$ parameter $%
I\rightarrow I_{ef}=I+\delta I_{ef}$ is determined by ``including'' Im$%
\Sigma _{{\bf k}}^{(3)}(E)$ into Im$\Sigma _{{\bf k}}^{(2)}(E)$ and is given
by 
\begin{equation}
\delta I_{ef}=-\frac N2I^2\int_{-\infty }^\infty d\omega \sum_{{\bf q}}{\cal %
J}_{{\bf q}}(\omega )\frac{n_{{\bf k-q}}}{E-t_{{\bf k-q}}-\omega }
\label{dipm}
\end{equation}
(here and hereafter we have to put $E=E_F=0,$ $k=k_F$ in the expressions for 
$I_{ef}$). To concretize the form of spin dynamics in the paramagnetic phase
we use below the spin diffusion approximation 
\begin{equation}
{\cal J}_{{\bf q}}(\omega )=\frac 1\pi \frac{{\cal D}q^2}{\omega ^2+({\cal D}%
q^2)^2}  \label{JD}
\end{equation}
(${\cal D}$ is the spin diffusion constant) which is correct at small ${\bf %
q,}\omega $ and is reasonable in a general case. Then we derive 
\begin{equation}
\delta I_{ef}=-\frac N2I^2\text{Re}\sum_{{\bf q}}\frac{n_{{\bf k-q}}}{E-t_{%
{\bf k-q}}-i{\cal D}q^2}
\end{equation}

To calculate the correction to the effective magnetic moment we treat the
static magnetic susceptibility 
\begin{equation}
\chi =(S^z,S^z)\equiv \int_0^{1/T}d\lambda \langle \exp (\lambda H)S^z\exp
(-\lambda H)S^z\rangle  \label{hi}
\end{equation}
Expanding to second order in $I$ we derive (cf. Ref.\cite{IKFTT}) 
\begin{eqnarray}
\chi &=&\,\overline{S}_{ef}^2/3T,\,\,\,\overline{S}_{ef}^2=S(S+1)[1-L]
\label{Sef} \\
L &=&2RI^2\int_{-\infty }^\infty d\omega \sum_{{\bf kq}}{\cal J}_{{\bf q}%
}(\omega )\frac{n_{{\bf k}}(1-n_{{\bf k-q}})}{(t_{{\bf k}}-t_{{\bf k-q}%
}-\omega )^2}  \nonumber
\end{eqnarray}
where we have introduced the notation 
\begin{equation}
R= 
{[l],\text{ model (\ref{sfe})} \atopwithdelims\{. N/2,\text{ model (\ref{CS})}}
\end{equation}
The spin-fluctuation frequency in the paramagnetic phase is determined from
the second moment of the spin Green's function 
\begin{equation}
\omega _{{\bf q}}^2=(\stackrel{.}{S}_{-{\bf q}}^z,\stackrel{.}{S}_{{\bf q}%
}^z)/(S_{-{\bf q}}^z,S_{{\bf q}}^z)
\end{equation}
To second order in $I$ we derive (cf.\cite{IKZ1,IKJP92} ) 
\begin{equation}
(\omega _{{\bf q}}^2)_0=\frac 43S(S+1)\sum_{{\bf p}}(J_{{\bf q-p}}-J_{{\bf p}%
})^2
\end{equation}
\begin{equation}
\delta \omega _{{\bf q}}^2/\omega _{{\bf q}}^2=(1-\alpha _{{\bf q}})\delta \,%
\overline{S}_{ef}^2/\,\overline{S}_{ef}^2=-(1-\alpha _{{\bf q}})L
\label{dwpm}
\end{equation}
where he have taken into account spin dynamics by analogy with (\ref{Sef}).
Passing into real space yields for the quantity $\alpha _{{\bf q}}$ 
\begin{equation}
\alpha _{{\bf q}}=\sum_{{\bf R}}J_{{\bf R}}^2\left( \frac{\sin k_FR}{k_FR}%
\right) ^2[1-\cos {\bf qR]/}\sum_{{\bf R}}J_{{\bf R}}^2[1-\cos {\bf qR]}
\label{Alpq}
\end{equation}
In the approximation of nearest neighbors at the distance $d,$ $\alpha $
does not depend on $q$: 
\begin{equation}
\alpha _{{\bf q}}=\alpha =\langle e^{i{\bf kR}}\rangle _{t_{{\bf k}%
}=E_F}=\left( \frac{\sin k_Fd}{k_Fd}\right) ^2  \label{ALp}
\end{equation}
so that we may use a single renormalization parameter, rather than the whole
function of {\bf q}.

\section{Effective interaction in magnetically ordered phases}

Now we investigate the renormalization of the $s-f$ interaction in FM and
AFM phases. First we treat the model (\ref{sfe}). For a ferromagnet the
electron spectrum possesses the spin splitting, $E_{{\bf k}\sigma }=t_{{\bf k%
}}-\sigma [l]IS.$ The second-order correction to $I_{ef}$ is determined by
the corresponding electron self-energies: 
\begin{equation}
\delta I_{ef}=-[\Sigma _{{\bf k\uparrow }}^{FM}(E)-\Sigma _{{\bf k\downarrow 
}}^{FM}(E)]/(2S[l])  \label{Ieffm}
\end{equation}
with 
\begin{eqnarray}
\Sigma _{{\bf k\uparrow }}^{FM}(E) &=&2RI^2S\sum_{{\bf q}}\frac{n_{{\bf k-q}}%
}{E-t_{{\bf k-q}}+\omega _{{\bf q}}^{FM}},\,  \nonumber \\
\Sigma _{{\bf k\downarrow }}^{FM}(E) &=&2RI^2S\sum_{{\bf q}}\frac{1-n_{{\bf %
k-q}}}{E-t_{{\bf k-q}}-\omega _{{\bf q}}^{FM}}  \label{sigfm}
\end{eqnarray}

For an antiferromagnet the electron spectrum contains the AFM gap, 
\begin{equation}
E_{{\bf k}}=\frac 12(t_{{\bf k}}+t_{{\bf k+Q}})\pm [\frac 14(t_{{\bf k}}-t_{%
{\bf k+Q}})^2+([l]IS)^2]^{1/2}  \label{afgap}
\end{equation}
The renormalization of $I$ is obtained from the second-order correction to
the anomalous Green's function 
\[
\langle \langle c_{{\bf k}\sigma }^{}|c_{{\bf k+Q}\sigma }^{\dagger }\rangle
\rangle _E=-\sigma \frac{[l]IS-\Sigma _{{\bf k,k+Q}}^{AFM}(E)}{(E-t_{{\bf k}%
})(E-t_{{\bf k+Q}})} 
\]
so that 
\begin{equation}
\delta I_{ef}=-\Sigma _{{\bf k,k+Q}}^{AFM}(E)/(S[l])
\end{equation}
The calculation of the off-diagonal self-energy gives 
\begin{equation}
\Sigma _{{\bf k,k+Q}}^{AFM}(E)=2RI^2S\sum_{{\bf q}}\frac{n_{{\bf k-q}}(E-t_{%
{\bf k-q}})}{(E-t_{{\bf k-q}})^2-(\omega _{{\bf q}}^{AFM})^2}  \label{sigafm}
\end{equation}

To calculate the corrections to $I_{ef}$ in the Coqblin-Schrieffer model we
consider the Green's function 
\begin{equation}
\sum_{M>0}(\langle \langle c_{{\bf k}M}^{}|c_{{\bf k+Q}M}^{\dagger }\rangle
\rangle _E-\langle \langle c_{{\bf k,-}M}^{}|c_{{\bf k+Q,-}M}^{\dagger
}\rangle \rangle _E)
\end{equation}
which determines the ``magnetization'' (FM case, ${\bf Q}=0$) or ``staggered
magnetization'' (AFM case) of conduction electrons.

For a ferromagnet the mean-field electron spectrum reads $E_{{\bf k}M}=t_{%
{\bf k}}-I\delta _{MS}$. Calculating the second-order corrections we derive 
\begin{equation}
\delta I_{ef}=-2[(N-1)\Sigma _{{\bf k\uparrow }}^{FM}(E)-\Sigma _{{\bf %
k\downarrow }}^{FM}(E)]/N
\end{equation}
(remember that in the Coqblin-Schrieffer model the self-energies should be
substituted at $S=1/2)$.

In the AFM1 state the mean-field electron spectrum for $M=S,-S$ is given by (%
\ref{afgap}) with the replacement $t_{{\bf k}}\rightarrow t_{{\bf k}}-I$.
For other $M$ the spectrum is unrenormalized. The renormalization of $I$ in
such a situation contains contributions of both FM and AFM types: 
\begin{equation}
\delta I_{ef}=-2[(N-2)\Sigma _{{\bf k\uparrow }}^{FM}(E,\omega _{{\bf q}%
}^{FM}\rightarrow \omega _{{\bf q}}^{(f)})+\Sigma _{{\bf k,k+Q}%
}^{AFM}(E,\omega _{{\bf q}}^{AFM}\rightarrow \omega _{{\bf q}}^{(a)})]/N
\end{equation}
In the case AFM2 the electron spectrum is given by (\ref{afgap}) with $%
S\rightarrow K/2,\,t_{{\bf k}}\rightarrow t_{{\bf k}}-IK.$ Besides that, for
odd $N$ there exists a branch of spectrum with $M=0,$ which is weakly
renormalized due to smallness of $|J_0^{(2)}|$. Then we obtain for even $N$%
\begin{equation}
\delta I_{ef}=-2\Sigma _{{\bf k,k+Q}}^{AFM}(E,\omega _{{\bf q}%
}^{AFM}\rightarrow \omega _{{\bf q}}^{(a)})
\end{equation}
For odd $N$ the contribution of the mode (\ref{Om}) occurs: 
\begin{eqnarray}
\delta I_{ef} &=&-2\frac{N-1}N\Sigma _{{\bf k,k+Q}}^{AFM}(E,\omega _{{\bf q}%
}^{AFM}\rightarrow \omega _{{\bf q}}^{(a)})  \nonumber \\
&&-[(K-n_0)\Sigma _{{\bf k,k+Q}}^{\Omega +}(E)+n_0\Sigma _{{\bf k,k+Q}%
}^{\Omega -}(E)]/K
\end{eqnarray}
where 
\begin{equation}
\Sigma _{{\bf k,k+Q}}^{\Omega \pm }(E)=I^2\sum_{{\bf q}}\frac{n_{{\bf k-q}%
}(E-t_{{\bf k-q}})}{\Omega _{{\bf q}}^{\pm }(E-t_{{\bf k-q}})}
\end{equation}

\section{Renormalizations of ordered moment and magnon frequency}

To investigate the magnon spectrum of an antiferromagnet in the model (\ref
{sfe}), we calculate the retarded Green's function of spin deviation
operators in the local coordinate system 
\begin{equation}
\Gamma _{{\bf q}}(\omega )=\langle \langle b_{{\bf q}}|b_{{\bf q}}^{\dagger
}\rangle \rangle _\omega ,\,\bar \Gamma _{{\bf q}}(\omega )=\langle \langle
b_{-{\bf q}}^{\dagger }|b_{{\bf q}}^{\dagger }\rangle \rangle _\omega
\end{equation}
Writing down the equation of motion we derive (cf. the calculations for $l=0$
\cite{IK96}) 
\begin{eqnarray}
\Gamma _{{\bf q}}(\omega ) &=&\frac{\omega +C_{{\bf q-}\omega }}{(\omega -C_{%
{\bf q}\omega })(\omega +C_{{\bf q-}\omega })+D_{{\bf q}\omega }^2}
\label{FGC} \\
\bar \Gamma _{{\bf q}}(\omega ) &=&\frac{D_{{\bf q}\omega }}{(\omega -C_{%
{\bf q}\omega })(\omega +C_{{\bf q-}\omega })+D_{{\bf q}\omega }^2}
\label{FGD}
\end{eqnarray}
where 
\begin{eqnarray}
C_{{\bf q}\omega } &=&S(J_{{\bf Q+q},\omega }^{tot}+J_{{\bf q}\omega
}^{tot}-2J_{{\bf Q}0}^{tot})+[l]\sum_{{\bf p}}[C_{{\bf p}}\Phi _{{\bf pq}%
\omega }^{AFM}  \nonumber \\
&&\ \ \ \ \ \ \ \ \ \ \ \ -(C_{{\bf p}}-D_{{\bf p}})\Phi _{{\bf p}%
00}^{AFM}+\phi _{{\bf pq}\omega }^{+}+\phi _{{\bf pq}\omega }^{-}]
\label{CD} \\
&&+\sum_{{\bf p}}[(2J_{{\bf Q}}+2J_{{\bf q-p}}-2J_{{\bf p}}-J_{{\bf Q+q}}-J_{%
{\bf q}})\langle b_{{\bf p}}^{\dagger }b_{{\bf p}}\rangle -2J_{{\bf p}%
}\langle b_{-{\bf p}}b_{{\bf p}}\rangle ]  \nonumber \\
D_{{\bf q}\omega } &=&D_{{\bf q-}\omega }=S(J_{{\bf q}\omega }^{tot}-J_{{\bf %
Q+q},\omega }^{tot})+[l]\sum_{{\bf p}}D_{{\bf p}}\Phi _{{\bf pq}\omega
}^{AFM}  \nonumber \\
&&+\sum_{{\bf p}}[(J_{{\bf Q+q}}-J_{{\bf q}})\langle b_{{\bf p}}^{\dagger
}b_{{\bf p}}\rangle -2J_{{\bf q-p}}\langle b_{-{\bf p}}b_{{\bf p}}\rangle ] 
\nonumber
\end{eqnarray}
The $s-f$ exchange contributions of the first order in $1/2S$ correspond to
the RKKY approximation 
\begin{equation}
J_{{\bf q}\omega }^{tot}=J_{{\bf q}}+I^2[l]\sum_{{\bf k}}\frac{n_{{\bf k}%
}-n_{{\bf k-q}}}{\omega +t_{{\bf k}}-t_{{\bf k-q}}}  \label{RKKY}
\end{equation}
the second summand in (\ref{RKKY}) being the $\omega $-dependent RKKY
indirect exchange interaction. The function $\Phi $, which determines the
second-order corrections, is given by 
\begin{eqnarray}
\Phi _{{\bf pq}\omega }^{AFM} &=&(\phi _{{\bf pq}\omega }^{+}-\phi _{{\bf pq}%
\omega }^{-})/\omega _{{\bf p}},  \label{Fi} \\
\phi _{{\bf pq}\omega }^{\pm } &=&I^2\sum_{{\bf k}}\frac{n_{{\bf k}}(1-n_{%
{\bf k+p-q}})+N(\pm \omega _{{\bf p}})(n_{{\bf k}}-n_{{\bf k+p-q}})}{\omega
+t_{{\bf k}}-t_{{\bf k+p-q}}\mp \omega _{{\bf p}}}
\end{eqnarray}
where $N(\omega )$ is the Bose function (note that $\phi _{{\bf pq}\omega
}^{+}=-\phi _{{\bf pq}-\omega }^{-}),$ $\omega _{{\bf p}}\,$is the magnon
frequency to zeroth order in $I\,$ and $1/2S$, given by (\ref{waf}), the
terms in (\ref{CD}) which contain the spin-deviation correlation functions
describe the magnon anharmonicity. The expression (\ref{FGC}) is valid also
for a ferromagnet (${\bf Q}=0$) provided that $q$ is not too small, cf. \cite
{Aus}; such an approximation is sufficient to treat the Kondo divergences.

We have to take into account singular $s-f$ contributions to the averages in
(\ref{CD}). These are due to the zero-point magnon damping and are obtained
by using the spectral representation for the Green's functions (\ref{FGC}), (%
\ref{FGD}) in the RKKY approximation \cite{IKFTT}. Since Im$J_{{\bf q}\omega
}^{tot}\sim \omega $ ($|\omega |\ll E_F)$ the corresponding integral over
frequency contains logarithmic Kondo-like divergences smeared by spin
dynamics (note that scattering processes correction to the damping described
by the function (\ref{Fi}) do not contribute the logarithmic terms). We
derive 
\begin{equation}
\delta \left\{ 
\begin{tabular}{l}
$\langle b_{{\bf q}}^{\dagger }b_{{\bf q}}\rangle $ \\ 
$\langle b_{-{\bf q}}b_{{\bf q}}\rangle $%
\end{tabular}
\right\} =\frac S2[l](\Phi _{{\bf q}00}^{AFM}\pm \Phi _{{\bf q+Q}00}^{AFM})
\label{bbaf}
\end{equation}
for an antiferromagnet and 
\begin{equation}
\delta \langle b_{{\bf q}}^{\dagger }b_{{\bf q}}\rangle =S[l]\Phi _{{\bf q}%
00}^{FM}  \label{bbf}
\end{equation}
with 
\begin{equation}
\Phi _{{\bf pq}\omega }^{FM}=I^2\sum_{{\bf k}}\frac{n_{{\bf k}}(1-n_{{\bf %
k+p-q}})+N(\omega _{{\bf p}})(n_{{\bf k}}-n_{{\bf k+p-q}})}{(\omega +t_{{\bf %
k}}-t_{{\bf k+p-q}}-\omega _{{\bf p}})^2}  \label{Fifm}
\end{equation}
for a ferromagnet. The expressions (\ref{bbaf}), (\ref{bbf}) determine also
the singular correction to the (sublattice) magnetization 
\begin{equation}
\delta \bar S/S=-\frac 1S\sum_{{\bf q}}\delta \langle b_{{\bf q}}^{\dagger
}b_{{\bf q}}\rangle =-[l]\sum_{{\bf q}}\Phi _{{\bf q}00}^{FM,AFM}
\label{Ssf}
\end{equation}
Collecting all the singular $s-d$ contributions to the pole of (\ref{FGC})
and taking into account the relation 
\begin{equation}
\sum_{{\bf p}}J_{{\bf p+k}}\Phi _{{\bf pq}\omega }\simeq \sum_{{\bf p}}J_{%
{\bf p+k-q}}\Phi _{{\bf p}00}\sim I^2\langle J_{{\bf k-q}}\rangle _{t_{{\bf k%
}}=t_{{\bf k}}=0}\ln \frac D{\overline{\omega }}
\end{equation}
which holds to logarithmic accuracy, we derive for the singular correction 
\begin{eqnarray}
\delta (\omega _{{\bf q}}^{AFM})^2 &\equiv &2\omega _{{\bf q}}\delta \omega
_{{\bf q}}^{AFM}\   \label{dw2} \\
\ &=&-R\sum_{{\bf p}}[2\omega _{{\bf q}}^2+4S^2(J_{{\bf Q+q}}+J_{{\bf q}%
}-2J_{{\bf Q}})(J_{{\bf p}}+J_{{\bf Q+p}}-J_{{\bf Q+q-p}}-J_{{\bf q-p}%
})]\Phi _{{\bf p}00}^{AFM}  \nonumber
\end{eqnarray}
with $R=[l].$ In the case of a ferromagnet (${\bf Q=}0$) the term $\phi _{%
{\bf pq}\omega }^{+}+\phi _{{\bf pq}\omega }^{-}$ (which is odd in $\omega $%
) yields a contribution to the pole of (\ref{FGC}) and we have 
\begin{eqnarray}
\delta \omega _{{\bf q}}^{FM}(\omega ) &=&-2RS\sum_{{\bf p}}(2J_{{\bf p}%
}-2J_{{\bf q-p}}+J_{{\bf q}}-J_0+\omega /2S)\Phi _{{\bf p}00}^{FM}
\label{dwf} \\
\delta \omega _{{\bf q}}^{FM} &=&\delta \omega _{{\bf q}}^{FM}(\omega _{{\bf %
q}})=-4RS\sum_{{\bf p}}(J_{{\bf p}}+J_{{\bf q}}-J_{{\bf q-p}}-J_0)\Phi _{%
{\bf p}00}^{FM}  \label{dwfp}
\end{eqnarray}
This expression may be represented as 
\begin{equation}
\delta \omega _{{\bf q}}/\omega _{{\bf q}}=2(1-\widetilde{\alpha }_{{\bf q}%
})\delta \overline{S}/S  \label{dwfm}
\end{equation}
with 
\begin{equation}
\widetilde{\alpha }_{{\bf q}}=\sum_{{\bf R}}J_{{\bf R}}\left| \langle e^{i%
{\bf kR}}\rangle _{t_{{\bf k}}=E_F}^2\right| ^2[1-\cos {\bf qR]/}\sum_{{\bf R%
}}J_{{\bf R}}[1-\cos {\bf qR]}  \label{Alpqf}
\end{equation}

For an antiferromagnet in the nearest-neighbor approximation ($J_{{\bf Q+q}%
}=-J_{{\bf q}}$) we obtain from (\ref{dw2}) 
\begin{equation}
\delta \omega _{{\bf q}}/\omega _{{\bf q}}=\delta \overline{S}/S
\label{dwafm}
\end{equation}
Note that the results (\ref{dwfm}),(\ref{dwafm}) differ from those of Ref.%
\cite{IKJP92} since only corrections arising from the static correlation
functions were taken into account in that paper.

The calculations of the magnon spectrum in the Coqblin-Schrieffer model for
FM and AFM1 cases are performed in a similar way by calculating the Green's
functions 
\begin{equation}
\Gamma _{{\bf q}}^{MM^{\prime }}(\omega )=\langle \langle \widehat{X}_{{\bf q%
}}^{M^{\prime }M}|\widehat{X}_{-{\bf q}}^{MM^{\prime }}\rangle \rangle
_\omega ,\,\bar \Gamma _{{\bf q}}^{MM^{\prime }}(\omega )=\langle \langle 
\widehat{X}_{{\bf q}}^{-M^{\prime },-M}|\widehat{X}_{-{\bf q}}^{MM^{\prime
}}\rangle \rangle _\omega  \label{Gamcs}
\end{equation}
for $M^{\prime }=S$. The results differ from (\ref{FGC}),(\ref{FGD}) by the
replacement $[l]\rightarrow 1$ in (\ref{RKKY}) and $[l]\rightarrow N/2$ in (%
\ref{CD}).

According to (\ref{SX}), the magnetization of a ferromagnet is determined by 
\begin{equation}
\bar S/S=1-Nn_{-},\,n_{-}=\sum_{{\bf q}}\langle X_{-{\bf q}}^{MS}X_{{\bf q}%
}^{SM}\rangle
\end{equation}
where the average in the right-hand side does not depend on $M$ for $M<S$.
Then we obtain 
\begin{equation}
\delta \bar S/S=-\frac N2\sum_{{\bf q}}\Phi _{{\bf q}00}^{FM},\,\delta
\omega _{{\bf q}}=\frac N2\delta \omega _{{\bf q}}^{FM}  \label{STF}
\end{equation}

The sublattice magnetization in the AFM1 case is given by 
\begin{equation}
\bar S/S=1-(N-2)n_{-}-2n_{-S}  \label{SAF1}
\end{equation}
with 
\begin{eqnarray}
n_{-} &=&\sum_{{\bf q}}\langle \hat X_{-{\bf q}}^{MS}\hat X_{{\bf q}%
}^{SM}\rangle \,(-S<M<S), \\
\,\,n_{-S} &=&\sum_{{\bf q}}\langle \hat X_{-{\bf q}}^{-SS}\hat X_{{\bf q}%
}^{S,-S}\rangle  \nonumber
\end{eqnarray}
After cumbersome calculations we derive 
\begin{equation}
\delta \bar S/S=-\frac{N-2}2\sum_{{\bf q}}\Phi _{{\bf q}00}^{FM}(\omega _{%
{\bf p}}^{FM}\rightarrow \omega _{{\bf p}}^{(f)})-\sum_{{\bf q}}\Phi _{{\bf q%
}00}^{AFM}(\omega _{{\bf p}}^{AFM}\rightarrow \omega _{{\bf p}}^{(a)})
\label{STAF1}
\end{equation}
\begin{eqnarray}
\delta \omega _{{\bf q}}^{(f)} &=&\frac{N-2}N\delta \omega _{{\bf q}%
}^{FM}(J_{{\bf p}}\rightarrow J_{{\bf p}}^{(2)})+\delta (\omega _{{\bf q}%
}^{AFM})^2/(NC_{{\bf q}}),  \label{dwAF1} \\
\delta (\omega _{{\bf q}}^{(a)})^2 &=&\frac 2N\delta (\omega _{{\bf q}%
}^{AFM})^2(\omega _{{\bf p}}^{AFM}\rightarrow \omega _{{\bf p}}^{(a)})+2%
\frac{N-2}NC_{{\bf q}}\delta \omega _{{\bf q}}^{FM}(\omega =\omega _{{\bf q}%
}^{(a)},\,J_{{\bf p}}\rightarrow J_{{\bf p}}^{(2)})  \nonumber
\end{eqnarray}
with $R=N/2$ in (\ref{dwf}), (\ref{dw2}).

In the AFM2 case we have to put in (\ref{Gamcs}) $M<0,\,M^{\prime }>0.$ Then
we have 
\begin{equation}
\bar S=(\bar S)_0- 
{(S+1/2)^2n_{-},\,\,N\text{ even} \atopwithdelims\{. S(S+1)n_{-}+(S+1)\delta n_0,\,N\text{ odd}}
\label{ST}
\end{equation}
where $(\bar S)_0$ is given by\ (\ref{S0}), the average 
\[
n_{-}=\sum_{{\bf q}}\langle \hat X_{-{\bf q}}^{M,M^{\prime }}\hat X_{{\bf q}%
}^{M^{\prime },M}\rangle 
\]
does not depend on $M,M^{\prime }$ for $M^{\prime }>0,\,M<0,$ and $\,\delta
n_0$ is the fluctuation correction to $n_0.$ Restricting ourselves for
simplicity to the case of even $N,$ which corresponds to a realistic
situation for $f$-ions, we obtain 
\begin{equation}
\delta \bar S/S=-\sum_{{\bf q}}\Phi _{{\bf q}00}^{AFM}(\omega _{{\bf p}%
}^{AFM}\rightarrow \omega _{{\bf p}}^{(a)})  \label{dSAF2}
\end{equation}
\begin{equation}
\delta \omega _{{\bf q}}^{(a)}=\frac 2N\delta \omega _{{\bf q}}^{AFM}(\omega
_{{\bf p}}^{AFM}\rightarrow \omega _{{\bf p}}^{(a)})  \label{dwAF2}
\end{equation}
Thus the corrections to the sublattice magnetization and magnon frequency do
not contain the factor of $N$ in this case.

\section{Scaling equations}

Using the results of previous sections we can write down the system of
scaling equations in the case of the Kondo lattice for various magnetic
phases. We use the ``poor man scaling'' approach \cite{And}. In this method
one considers the dependence of effective (renormalized) model parameters on
the cutoff parameter $C$ which occurs at picking out the Kondo singular
terms. To find the equation for $I_{ef}$ we calculate in the sums in (\ref
{dipm}), (\ref{sigfm}), (\ref{sigafm}) the contribution of intermediate
electron states near the Fermi level with $C<t_{{\bf k+q}}<C+\delta C.$ Then
we obtain for the model (\ref{sfe}), and for the PM, FM and AFM2 phases in
the Coqblin-Schrieffer model 
\begin{equation}
\delta I_{ef}(C)=N\rho I^2\eta (-\frac{\overline{\omega }}C)\delta C/C
\label{ief}
\end{equation}
where $\overline{\omega }$ is a characteristic spin fluctuation energy, $N=2$
for the $s-f$ model, $\eta (x)$ is the scaling function which satisfies the
condition $\eta (0)=1$ (this guarantees the correct one-impurity limit, cf.
Ref.\cite{And}). For the paramagnetic phase we obtain 
\begin{equation}
\eta ^{PM}(\frac{\overline{\omega }}C)=\text{Re}\int_{-\infty }^\infty
d\omega \langle {\cal J}_{{\bf k-k}^{\prime }}(\omega )\rangle
_{t_k=t_{k^{\prime }}=E_F}\frac 1{1-(\omega +i0)/C}  \label{etpm}
\end{equation}
For FM and AFM phases we have 
\begin{eqnarray}
\eta ^{FM}(\frac{\overline{\omega }}C) &\equiv &\frac 1N[(N-1)\eta
_{\uparrow }(\frac{\overline{\omega }}C)+\eta _{\downarrow }(\frac{\overline{%
\omega }}C)]  \label{etfm} \\
\eta _{\uparrow ,\downarrow }^{FM}(-\frac{\overline{\omega }}C)\ &=&\langle
(1\mp \omega _{{\bf k-k}^{\prime }}/C)^{-1}\rangle _{t_k=t_{k^{\prime }}=E_F}
\nonumber
\end{eqnarray}
\begin{equation}
\eta ^{AFM}(-\frac{\overline{\omega }}C)=\langle (1-\omega _{{\bf k-k}%
^{\prime }}^2/C^2)^{-1}\rangle _{t_k=t_{k^{\prime }}=E_F}  \label{etafm}
\end{equation}
Using the spin diffusion approximation (\ref{JD}) in (\ref{etpm}) and the
approximations $\omega _{{\bf q}}^{FM}\sim q^2,$ $\omega _{{\bf q}%
}^{AFM}\sim q$ (which are justified, e.g., at small $k_F$)$,$ we get

\begin{eqnarray}
\eta ^{PM}(x) &=&x^{-1}\arctan x  \label{eta} \\
\eta _{\uparrow ,\downarrow }^{FM}(x) &=&\pm x^{-1}\ln |1\pm x|  \nonumber \\
\eta ^{AFM}(x) &=& 
{-x^{-2}\ln |1-x^2|,\,d=3 \atopwithdelims\{. (1-x^2)^{-1/2}\theta (1-x^2),\,d=2}
\nonumber
\end{eqnarray}
where $\overline{\omega }=4{\cal D}k_F^2$ for a paramagnet, $\overline{%
\omega }=\omega _{2k_F}$ for FM and AFM cases, $\theta (x)$ is the step
function. One can see that the scaling functions for the ordered phases
contain singularities at $x=1.$ Presence of this singularity is a general
property which does not depend on the spectrum model. The functions $\eta
_{\downarrow }^{FM}(x)$ and $\eta ^{AFM}(x)$ ($d=3)$ change their sign at $%
x=2$ and $x=\sqrt{2}$ respectively. For $d=2$ the function $\eta ^{AFM}(x)$
vanishes discontinuosly at $x>1$, but a smooth contribution occurs for more
realistic models of magnon spectrum.

The singularities are retained if we introduce the gap in the spin-wave
spectrum. For a ferromagnet $\omega _{{\bf q}}$ is replaced by $\omega
_0+\omega _{{\bf q}}$. Then we have, besides the singularity at $%
x\rightarrow 1,$ the second singularity at $x>1$: 
\begin{equation}
\widetilde{\eta }_{\uparrow ,\downarrow }^{FM}(x)=\pm \frac{\omega _0+\omega
_{ex}}{2x\omega _{ex}}\ln \left| \frac{1\pm x}{1\pm \omega _0x/(\omega
_0+\omega _{ex})}\right|  \label{etaanf}
\end{equation}
where $\omega _{ex}=\max \omega _{{\bf q}}.$ In the AFM case $\omega _{{\bf q%
}}^2$ is replaced by $\omega _0^2+\omega _{{\bf q}}^2$ and we derive 
\begin{equation}
\widetilde{\eta }^{AFM}(x)=\frac{\omega _0^2+\omega _{ex}^2}{\omega
_0^2(1-x^2)+\omega _{ex}^2}\eta ^{AFM}\left( \frac{x\omega _{ex}}{[\omega
_0^2(1-x^2)+\omega _{ex}^2]^{1/2}}\right)  \label{etaan}
\end{equation}
We shall see that the latter singularity may influence considerably the
behavior of the effective coupling constant. In the limit of strong magnetic
anisotropy $\omega _0/\omega _{ex}\rightarrow \infty $ the singularity at $%
x\rightarrow 1$ is very strong: 
\begin{equation}
\eta ^0(x)=(1-x^2)^{-1}
\end{equation}

When considering characteristics of localized-spin subsystem, the
lowest-order Kondo corrections originate from double integrals over both
electron and hole states (see (\ref{Sef}), (\ref{Fifm}), (\ref{Fi})). Then
we have to introduce two cutoff parameters $C_e$ and $C_h$ with $C_e+C_h=C$ (%
$C$ is the cutoff parameter for the electron-hole excitations). In the FM
case for the Coqblin-Schrieffer model we have $\delta C_h=-(N-1)\delta C_e$
due to the requirement of the number-of-particle conservation in
electron-hole transitions (there exists $N-1$ ``channels'' for electrons and
one ``channel'' for holes). For PM and AFM2 cases in the Coqblin-Schrieffer
model, as well as for all the cases in the $s-f$ model, the electron-hole
symmetry is not violated and we have $\delta C_e=-\delta C_h.$

Taking into account (\ref{Sef}), (\ref{STF}), (\ref{dSAF2}) we obtain 
\begin{equation}
\delta \overline{S}_{ef}(C)/S=V\rho \delta I_{ef}(C)=VN\rho ^2I^2\eta (-%
\frac{\overline{\omega }}C)\delta C/C  \label{sef}
\end{equation}
where $V=[l]$ for the $s-f$ model, $V=1$ for PM and FM phases in the
Coqblin-Schrieffer model and $V=2/N$ for the AFM2 phase. The
renormalizations of spin-wave frequencies are obtained in a similar way from
(\ref{dwpm}), (\ref{dwf}), (\ref{dw2}), (\ref{STF}), (\ref{dwAF2}), and are
given by 
\begin{equation}
\delta \overline{\omega }_{ef}(C)/\overline{\omega }=a\delta \overline{S}%
_{ef}(C)/S=aVN\rho ^2I^2\eta (-\frac{\overline{\omega }}C)\delta C/C
\label{wef}
\end{equation}
where, in the nearest-neighbor approximation, 
\begin{equation}
a=\left\{ 
\begin{array}{cc}
1-\alpha & \text{PM} \\ 
2(1-\alpha ) & \text{FM} \\ 
1 & \text{AFM}
\end{array}
\right.
\end{equation}
Introducing the dimensionless coupling constants 
\[
g_{ef}(C)=-N\rho I_{ef}(C),\,g=-N\rho I 
\]
and replacing in the right-hand parts of (\ref{ief}), (\ref{wef}), (\ref{sef}%
) $g\rightarrow g_{ef}(C),\,\overline{\omega }\rightarrow \overline{\omega }%
_{ef}(C)$ we obtain the system of scaling equation 
\begin{eqnarray}
\partial g_{ef}(C)/\partial C &=&-\Lambda  \label{gl} \\
\partial \ln \overline{\omega }_{ef}(C)/\partial C &=&aV\Lambda /N
\label{wl} \\
\partial \ln \overline{S}_{ef}(C)/\partial C &=&V\Lambda /N  \label{sl}
\end{eqnarray}
with 
\begin{equation}
\Lambda =\Lambda (C,\overline{\omega }_{ef}(C))=[g_{ef}^2(C)/C]\eta (-%
\overline{\omega }_{ef}(C)/C)
\end{equation}

As regards the AFM1 state in the Coqblin-Schrieffer model, its treatment in
a general case is a difficult and cumbersome problem. However, one can see
from (\ref{dwAF1}) that in the leading approximation in $1/N$ the scaling
equations coincide with those for the FM state with the replacement $J_{{\bf %
p}}\rightarrow J_{{\bf p}}^{(2)}.$

\section{Improved version of scaling equations: an account of dissipative
contributions to spin dynamics in ordered phases}

The transition from the dissipative spin dynamics, which is characteristic
for the PM phase, to the dynamics with well-defined spin-wave excitations
(ordered FM and AFM phases) results in occurrence of singularities in the
scaling function $\eta (x)$ at $x\rightarrow 1$ and in a decrease of the $%
g_c $ value. One may suppose that in the situation of strongly suppressed
saturation moment ($g$ is close to $g_c)$ the character of spin dynamics in
the ordered phases should change drastically. By the analogy with weak
itinerant magnets \cite{Mor} one may expect that for $\overline{S}\ll S$ a
considerable part of the localized-spin spectral density comes from the
branch cut of the spin Green's function rather than from the magnon pole.

In this section we shall demonstrate that this indeed takes place provided
that our approach is slightly modified. To this end we shall analyze the
structure of the spin Green's function with account of the singular Kondo
corrections in more detail.

First we consider the case of a ferromagnet. We obtain near the magnon pole 
\begin{equation}
\langle \langle S_{{\bf q}}^{+}|S_{-{\bf q}}^{-}\rangle \rangle _\omega =%
\frac{2\overline{S}Z_{{\bf q}}}{\omega -\omega _{{\bf q}}^{ef}}+\langle
\langle S_{{\bf q}}^{+}|S_{-{\bf q}}^{-}\rangle \rangle _\omega ^{incoh}
\label{gz}
\end{equation}
where the residue at the pole is determined by (\ref{dwf}): 
\begin{equation}
1/Z_{{\bf q}}-1=-\left( \frac{\partial \omega _{{\bf q}}(\omega )}{\partial
\omega }\right) _{\omega =\omega _{{\bf q}}}\simeq -[l]\sum_{{\bf p}}\Phi _{%
{\bf p}00}^{FM}
\end{equation}
Besides that, there exists the singular contribution which comes from the
incoherent (non-pole) part of the spin spectral density. To calculate the
renormalization of $g$ we use, instead of (\ref{sigfm}), the representation
of the electron self-energy in terms of the spectral density 
\begin{eqnarray}
\Sigma _{{\bf k\uparrow }}^{FM}(E) &=&-\frac 1\pi RI^2\int_{-\infty }^\infty
d\omega \sum_{{\bf q}}\frac{n_{{\bf k-q}}}{E-t_{{\bf k-q}}+\omega }\text{Im}%
\langle \langle S_{{\bf q}}^{+}|S_{-{\bf q}}^{-}\rangle \rangle _\omega 
\nonumber \\
\Sigma _{{\bf k\downarrow }}^{FM}(E) &=&-\frac 1\pi RI^2\int_{-\infty
}^\infty d\omega \sum_{{\bf q}}\frac{1-n_{{\bf k-q}}}{E-t_{{\bf k-q}}-\omega 
}\text{Im}\langle \langle S_{{\bf q}}^{+}|S_{-{\bf q}}^{-}\rangle \rangle
_\omega
\end{eqnarray}
Thus the magnon pole contribution to $g_{ef}$ is multiplied by $Z$, and the
incoherent contribution by $1-Z.$ The renormalizations of $\overline{\omega }%
_{ef}$ and $Z$ are obtained in a similar way. We derive in the
nearest-neighbor approximation 
\begin{eqnarray}
\partial g_{ef}(C)/\partial C &=&-\Lambda  \label{gll} \\
\partial \ln \overline{\omega }_{ef}(C)/\partial C &=&aV\Lambda /N
\label{wll} \\
\partial (1/Z)/\partial C &=&\partial \ln \overline{S}_{ef}(C)/\partial
C=V\Lambda /N  \label{zl}
\end{eqnarray}
where $a=2(1-\alpha ),\,N=2$%
\begin{equation}
\Lambda =[g_{ef}^2(C)/C][Z\eta _{coh}(-\overline{\omega }_{ef}(C)/C)+(1-Z)%
\eta _{incoh}(-\overline{\omega }_{ef}(C)/C)]  \label{l1}
\end{equation}
$\eta _{coh}=\eta ^{FM},$ and the function $\eta _{incoh}$ is, generally
speaking, unknown. For further estimations we may put $\eta _{incoh}=\eta
^{PM}.$

Although the account of the incoherent part does not modify strongly
numerical results (see Sect.10), the physical picture of magnetism changes
drastically. According to (\ref{zl}) we have 
\begin{equation}
\frac 1{Z(\xi )}=1+\ln \frac S{\overline{S}(\xi )}  \label{1/ZS}
\end{equation}
Consequently, the increase of magnetic moment owing to the Kondo screening
leads to a considerable suppression of magnon contributions to the spectral
density. Unlike the case of weak itinerant magnets, this suppression is
logarithmic.

In the case of an antiferromagnet the calculations are performed by taking
into account the expressions (\ref{FGC}), (\ref{FGD}). In the
nearest-neighbor approximation ($J_{{\bf Q+q}}=-J_{{\bf q}}$) we obtain 
\begin{equation}
\langle \langle S_{{\bf q}}^{+}+S_{{\bf q}}^{-}|S_{-{\bf q}}^{+}+S_{-{\bf q}%
}^{-}\rangle \rangle _\omega ^{coh}\simeq \frac{8S^2(J_0-J_{{\bf q}})}{%
\omega ^2-(\omega _{{\bf q}}^{ef})^2}(1-[l]\sum_{{\bf p}}\Phi _{{\bf p}%
00}^{AFM})  \label{FGAF}
\end{equation}
(the quantity (\ref{FGAF}) just determines the Kondo renormalization of
electron spectrum, cf. Ref.\cite{IKJP92}).Then the results for AFM differ
from those for FM by the replacement $a\rightarrow 1$ in the scaling
equations (\ref{gll})-(\ref{zl}) only.

\section{The scaling behavior in the large-$N$ Coqblin-Schrieffer model}

It is instructive to consider the limit $N\rightarrow \infty .$ (To avoid
misunderstanding, it should be noted that this limit with inclusion of spin
dynamics in the zeroth approximation differs somewhat from the
considerations of Refs.\cite{Bick,New}).Then the renormalizations of spin
dynamics and $\overline{S}_{ef}$ are absent, and the transition into the
non-magnetic Kondo-lattice state cannot be described. However, peculiarities
of the dependence $g_{ef}(C)$ for various types of magnetic ordering are
described qualitatively, explicit analytical expressions being obtained. We
have 
\begin{equation}
1/g_{ef}(C)-1/g=G(C)=-\int_{-D}^C\frac{dC^{\prime }}{C^{\prime }}\eta (-%
\frac{\overline{\omega }}{C^{\prime }})  \label{intG}
\end{equation}
where $D$ is the cutoff energy defined by $g_{ef}(-D)=g.$ Performing
integration we obtain 
\begin{eqnarray}
G^{PM}(C) &=&\frac 12\ln ((C^2+\overline{\omega }^2)/D^2)+\frac C{\overline{%
\omega }}\arctan (\frac{\overline{\omega }}C)-1  \label{1g} \\
G^{FM}(C) &=&\ln |C/D|-(C/\overline{\omega }-1)\ln |1-\overline{\omega }/C|+1
\label{g2g} \\
G^{AFM}(C) &=&\ln |C/D|-\frac 12[(C^2/\overline{\omega }^2-1)\ln |1-%
\overline{\omega }^2/C^2|-1],\,d=3  \label{3g} \\
G^{AFM}(C) &=&\theta (|C|-\overline{\omega })\ln (\frac 12(|C|+\sqrt{C^2-%
\overline{\omega }^2})/D)+\theta (\overline{\omega }-|C|)\ln (\overline{%
\omega }/2D),\,d=2  \label{4g} \\
G^0(C) &=&\frac 12\ln |(C^2-\overline{\omega }^2)/D^2|  \label{5g}
\end{eqnarray}

The dependences $1/g_{ef}(\xi =\ln |D/C|)$ are shown in Fig.1. The effective
coupling constant $g_{ef}(C)$ begins to deviate strongly from its
one-impurity behavior 
\begin{equation}
1/g_{ef}(C)=1/g-\ln |D/C|
\end{equation}
at $|C|\sim \overline{\omega }.$ One can see that at small $\overline{\omega 
}\ll |C|$ spin dynamics results in a decrease of $g_{ef}(C)$ for PM and FM
cases, and in an increase for AFM case and for the case of strong
anisotropy. For the dispersionless spin dynamics mode the singularity at $%
|C|=\overline{\omega }$ in (\ref{intG}) is non-integrable, and consequently $%
g_{ef}(C)$ diverges at this point for arbitrary small $g.$ However, an
account of a small dispersion or damping (the latter occurs in next orders
in $g,$ cf.\cite{IKloc}) results in eliminating this divergence. Thus this
case needs a more consistent consideration with account of higher-order
corrections, which will be performed elsewhere.

It should be noted that the equation (\ref{intG}) can be used even for small 
$N$ provided that $g$ is considerably smaller than the critical value $g_c$
(see the next Section). Besides that, the equation (\ref{intG}) works in PM
and FM phases provided that $k_F$ is small, so that, according to (\ref{ALp}%
), $\alpha \rightarrow 1.$ However, in the case of the ferromagnet we have
instead of (\ref{g2g}) 
\begin{eqnarray}
G^{FM}(C) &=&\ln |C/D|-\frac{N-1}N(C/\overline{\omega }-1)\ln |1-\overline{%
\omega }/C|  \label{g22g} \\
&&+\frac 1N(C/\overline{\omega }+1)\ln |1+\overline{\omega }/C|+1  \nonumber
\end{eqnarray}

In the 3D AFM case $1/g_{ef}(C)$ has a minimum with decreasing $|C|.$ For a
2D AFM, $1/g_{ef}$ has a square-root singularity at $|C|<\overline{\omega }$
and is constant at $|C|>\overline{\omega }$. As we shall see in the next
Section, these features retain at finite $N$. The minimum occurs also in the
2D AFM case provided that we introduce even small magnetic anisotropy. In
the 3D case, the anisotropy results in that the minimum becomes deeper. For
an anisotropic ferromagnet, the minimum is absent at $N\rightarrow \infty $
since only the contribution of $\eta _{\uparrow }$ survives. However, it is
present for finite $N$ due to contribution of $\eta _{\downarrow }$. The
picture for $N=2$ and $g$ well below $g_c$ (when the equation (\ref{intG})
works qualitatively) is shown in Fig.2. Note that, contrary to the case $%
N\rightarrow \infty ,$ for FM with $N=2$ spin dynamics results in a decrease
of $g_{ef}(C)$ at not too small $C.$

The boundary of the strong coupling region (the renormalized Kondo
temperature) is determined by $G(C=-T_K^{*})=-1/g.$ Of course, $T_K^{*}$
means here only some characteristic energy scale extrapolated from high
temperatures, and the detailed description of the ground state requires a
more detailed consideration. In the PM and FM phases spin dynamics
suppresses $T_K^{*}.$ On the other hand, in the AFM case spin dynamics at
not large $\overline{\omega }$ results in and increase of $T_K^{*}.$

Provided that the strong coupling regime does not occur, i.e. $g$ is smaller
than the critical value $g_c$, $g_{ef}(C\rightarrow 0)$ tends to a finite
value $g^{*}.$ To leading order in $\ln (D/\overline{\omega })$ we have 
\begin{equation}
1/g_c=\lambda \equiv \ln (D/\overline{\omega })
\end{equation}
However, an account of next-order terms results in an appreciable dependence
on the type of magnetic ordering and space dimensionality. For PM, FM and 2D
AFM phases the critical value $g_c$ is given by $1/g_c=-G(0)$. Then we
obtain 
\begin{equation}
1/g^{*}=1/g-1/g_c  \label{glin}
\end{equation}
with 
\begin{equation}
1/g_c=\lambda +\left\{ 
\begin{tabular}{ll}
1 & PM, FM \\ 
$\ln 2$ & 2D AFM
\end{tabular}
\right.  \label{g33}
\end{equation}
At the same time, in the 3D AFM case (and also in the anisotropic 2D case) $%
g_c$ is determined by the minimum of the function $G(C)$. Thus $g^{*}$
remains finite at $g\rightarrow g_c-0$: 
\begin{equation}
1/g^{*}=1/g-1/g_c+1/g_c^{*}  \label{gglin}
\end{equation}
where 
\begin{eqnarray}
1/g_c &=&\lambda +(1+\ln 2)/2,  \label{g44} \\
1/g_c^{*} &=&\Delta =(\ln 2)/2\simeq 0.35  \nonumber
\end{eqnarray}
$\Delta $ being the depth of the minimum.

For $C\rightarrow 0$ we have 
\begin{equation}
G(C)-G(0)\simeq \left\{ 
\begin{tabular}{ll}
$\pi |C|/2\overline{\omega }$ & PM \\ 
$|C/\overline{\omega }|\ln |\overline{\omega }/C|$ & FM \\ 
$-(C/\overline{\omega })^2\ln |\overline{\omega }/C|$ & 3D AFM
\end{tabular}
\right.  \label{G11}
\end{equation}
(however, $G^{FM}(C)-G^{FM}(0)\simeq (C/\overline{\omega })^2$ for the
function (\ref{g22g}) at $N=2$). As follows from (\ref{G11}), for $g$ $%
\rightarrow g_c+0$ we have in PM and FM cases to logarithmic accuracy 
\begin{equation}
T_K^{*}\simeq \overline{\omega }\frac{g-g_c}{g_c^2}\times 
{2/\pi ,\text{ PM } \atopwithdelims\{. \ln [(g-g_c)/g_c],\text{ FM}}
\label{TKG}
\end{equation}
On the other hand, in the AFM phases $T_K^{*}$ is finite at $g\rightarrow
g_c+0$: 
\begin{equation}
T_K^{*c}=\overline{\omega }\times 
{1,\text{ 2D AFM} \atopwithdelims\{. 1/\sqrt{2},\,\text{3D AFM}}
\label{t33}
\end{equation}

\section{Scaling behavior for finite $N$}

Some results of the large-$N$ limit hold in the finite-$N$ case too, but
there occur a number of new important features. To consider the general case
we write down the first integral of the system (\ref{gl}), (\ref{wl}) 
\begin{equation}
g_{ef}(C)+(N/Va)\ln \overline{\omega }_{ef}(C)=\text{const}  \label{int}
\end{equation}
which results in 
\begin{equation}
\overline{\omega }_{ef}(C)=\overline{\omega }\exp (-(aV/N)[g_{ef}(C)-g])
\label{w+g}
\end{equation}
As follows from (\ref{wl}), (\ref{sl}) 
\begin{equation}
\frac{\overline{S}_{ef}(C)}S=\left( \frac{\overline{\omega }_{ef}(C)}{%
\overline{\omega }}\right) ^{1/a}  \label{sw}
\end{equation}
Substituting (\ref{w+g}) into (\ref{gl}) we obtain 
\begin{equation}
\partial (1/g_{ef})/\partial \xi =-\Psi (\lambda +(aV/N)[g_{ef}-g]-\xi )
\label{gpsi}
\end{equation}
where 
\[
\Psi (\xi )=\eta (e^{-\xi }),\,\xi =\ln |D/C|,\,\lambda =\ln (D/\overline{%
\omega })\gg 1 
\]
Presence of $g_{ef}$ in the argument of the function $\Psi $ in (\ref{gpsi})
leads to drastic changes in the scaling behavior in comparison with the
large-$N$ limit. We describe below the renormalization process in various
cases.

If $g$ is sufficiently large, the increase of $g_{ef}$ will lead to that the
argument of $\Psi $ will never be small, so that the scaling behavior will
be essentially the one-impurity one. At smaller $g$ the function $%
1/g_{ef}(\xi )$ begins to deviate strongly from the one-impurity behavior $%
1/g_{ef}(\xi )=1/g-\xi $ starting from $\xi \simeq \xi _1$ where $\xi _1$ is
the minimal solution to the equation 
\begin{equation}
\lambda +(aV/N)[g_{ef}(\xi )-g]=\xi  \label{eq}
\end{equation}
If the argument remains negative with further increasing $\xi ,$ $%
1/g_{ef}(\xi )$ will decrease tending to the finite value $g^{*},$ the
derivative $\partial (1/g_{ef})/\partial \xi $ being exponentially small, so
that the situation is close to the large-$N$ case (Fig.1). However, $%
(aV/N)g_{ef}(\xi )$ can increase more rapidly than $\xi ,$ so that the
second solution to (\ref{eq}), $\xi _2$, will occur, and the argument of the
function $\Psi $ becomes positive again. Then $g_{ef}(\xi )$ will diverge at
some point $\xi ^{*}=\ln |D/T_K^{*}|.$ The divergence is described, unlike
the large-$N$ limit, by the law 
\begin{equation}
g_{ef}(\xi )\simeq 1/(\xi ^{*}-\xi )  \label{div}
\end{equation}
since $\eta (x\ll 1)=1.$ The behavior (\ref{div}) takes place starting from $%
\xi \simeq \xi _2.$

The dependences $g_{ef}(\xi )$ in the PM case at small $|g-g_c|$ are shown
in Fig.3. The behavior $g_{ef}(\xi )$ between $\xi _1$ and $\xi _2$ is
nearly linear, but is somewhat smeared since $\Psi (\xi )$ differs
considerably from the asymptotic values 0 and 1 in a rather large interval
of $\xi .$

The case of magnetically ordered phases has a number of peculiarities. Here
the singularity of the function $\Psi (\xi )$ at $\xi =0$ turns out to play
the crucial role. In particular, one can prove that $g_{ef}$ diverges at
some $\xi $ at arbitrarily small $g$ (i.e. $g_c=0$) unless the singularity
cutoff is introduced. Indeed, when approaching the singularity point with
increasing $\xi ,$ the derivative $\partial g_{ef}/\partial \xi $ rapidly
increases, and the argument of the function $\Psi $ in (\ref{gpsi})
inevitably starts to increase at some point $\xi _1.$ Thus the singularity
point cannot be crossed, and the argument of the function $\Psi $ is always
positive. Since $\Psi (\xi ^{\prime }>0)>1$, we have $\partial
(1/g_{ef})/\partial \xi <-1$ at arbitrary $\xi .$ Therefore the effective
coupling constant diverges at $\xi =\xi ^{*}<1/g$.

To make the value of $g_c$ finite one has to cut the singularity of the
scaling functions. This may be performed by introducing small imaginary
parts, i.e. replacing in (\ref{eta}) 
\begin{eqnarray}
&&\ \ln |1-x| 
\begin{array}{c}
\rightarrow
\end{array}
\frac 12\ln [(1-x)^2+\delta ^2],  \label{cut} \\
&&\ (1-x)^{-1/2}\theta (1-x) 
\begin{array}{c}
\rightarrow
\end{array}
\text{Re}(1-x+i\delta )^{-1/2}  \nonumber \\
\ &=&\{[(1-x)^2+\delta ^2]^{1/2}+1-x]/2\}^{1/2}/[(1-x)^2+\delta ^2]^{1/2} 
\nonumber
\end{eqnarray}
Then $\Psi $ becomes bounded from above: 
\begin{equation}
\Psi _{\max }=\eta _{\max }\simeq \eta (1)\simeq \left\{ 
\begin{tabular}{ll}
$\frac 12\ln \delta $ & FM \\ 
$\ln \delta $ & 3D AFM \\ 
$\frac 12\delta ^{-1/2}$ & 2D AFM
\end{tabular}
\right.
\end{equation}
Other way to cut the divergence in the 2D AFM case is to retain strictly the
proportionality to the step function 
\begin{equation}
(1-x)^{-1/2}\theta (1-x)\rightarrow \text{Re}(1-x+i\delta )^{-1/2}\theta
(1-x)  \label{cut1}
\end{equation}
so that $\Psi (\xi )$ cannot take small values.

The value of $\delta $ should be determined by the magnon damping at $q=|%
{\bf k-k}^{\prime }|\simeq 2k_F$ (see (\ref{etfm}), (\ref{etafm})). This
damping is due to both exchange and relativistic interactions. The damping
owing to exchange scattering by conduction electrons should be formally
neglected within the one-loop approximation, since this contains a more high
power of $I.$ The magnon-magnon interaction in the Heisenberg model gives
the damping at non-zero temperatures only. However, the relativistic (e.g.,
dipole) interactions give a damping at $T=0$ due to zero-point oscillations.
Hereafter we put in numerical calculations $\delta =1/100.$

The behavior of the solutions to (\ref{gpsi}) in magnetic phases for $g$
well below $g_c$ is similar to that in the large-$N$ limit (see Fig.2). In
particular, the function $1/g_{ef}(\xi )$ has a minimum in the 3D AFM case
with the same depth, and a similar situation occurs in the presence of
anisotropy. The presence of the minimum may result in non-monotonic
temperature dependences of physical quantities which are sensitive to the
Kondo screening, e.g., of the effective magnetic moment. These dependences
are obtained qualitatively by the replacement $|C|\rightarrow T$ in (\ref
{w+g}),(\ref{sw}). Of course, the standard monotonic spin-wave corrections
should be added to the Kondo contributions.

It is important that, except for a very narrow region near $g_c,$ at
approaching $g_c$ the value of $g^{*}$ becomes practically constant, $%
g^{*}\simeq g(\xi _1)=g_1.$ This value is estimated as 
\begin{equation}
(aV/N)(\partial g_{ef}/\partial \xi )_{\max }=(aV/N)g_1^2\Psi _{\max }=1
\label{g1e}
\end{equation}
On the other hand, we may estimate from the linear asymptotics $1/g_{ef}(\xi
)\simeq 1/g-\xi $ (which holds up to $\xi \simeq \xi _1\simeq \lambda
+(aV/N)g_1)$%
\begin{equation}
1/g_1\simeq 1/g-\xi _1  \label{g2e}
\end{equation}
Comparing (\ref{g2e}) and (\ref{g1e}) we derive the rough estimation 
\begin{eqnarray}
1/g_c &=&\lambda +(aV/N)g_1+1/g_1  \label{gce} \\
\ &=&\lambda +(N\Psi _{\max }/aV)^{-1/2}+(aV\Psi _{\max }/N)^{1/2}  \nonumber
\end{eqnarray}
The value of $1/g_2$ can be also estimated from the linear asymptotics (\ref
{div}): 
\begin{equation}
1/g_2\simeq \xi ^{*}-\xi _2  \label{g66}
\end{equation}
Owing to the singularity, $\xi _2$ is close to $\xi _1$ except for very
small $|g-g_c|.$ In the latter case the behavior $g(\xi _1<\xi <\xi _2)$ is
practically linear, 
\begin{equation}
(aV/N)g_{ef}(\xi _1<\xi <\xi _2)\simeq \xi -\lambda  \label{linn}
\end{equation}
(this behavior is discussed in detail in the next sections). This is
explained by that the argument of the function $\Psi $ in (\ref{gpsi})
should be nearly zero. The behavior at $\xi >\xi _2$ is described by (\ref
{div}). We may estimate from (\ref{g66}) 
\begin{equation}
\xi ^{*}\simeq 1/g+1/g_2-1/g_1
\end{equation}

For the cutoff (\ref{cut1}) with $\eta (x>1)=0$ in the 2D AFM case (and in
similar situations) $\xi ^{*}$ tends to a finite limit $\xi _c^{*}$ at $%
g\rightarrow g_c+0.$ Indeed, as follows from the form of the scaling
function (\ref{eta}), one has up to the divergence point $\Psi (\xi )>1,$ so
that $\xi ^{*}<1/g<1/g_c.$ In such a situation $g^{*}$ turns out to be also
finite at $g\rightarrow g_c-0$, and the character of approaching $g_c$ is
quite different from that in the large-$N$ limit. With increasing $g,$ the
position of the $1/g_{ef}(\xi )$ singularity point is shifted to right due
to rapid increase of $g_{ef}$ in the argument of the function $\Psi $ in (%
\ref{gpsi}). The shift should stop at $\xi _1<\xi _c^{*}.$ This takes place
just at $g\rightarrow g_c-0.$ Thus $1/g_{ef}(\xi )$ should vanish
discontinuously at $g=g_c.$ The ``maximum'' value of $g_{ef}(\xi ),$ $%
g_{ef}^{\max }=g_c^{*},$ is estimated from 
\begin{equation}
\lambda +(aV/N)(g_{ef}^{\max }-g_c)-\xi _1=0  \label{ksic}
\end{equation}
Since for $g>g_c$ the decrease $1/g_{ef}(\xi )$ at $\xi >\xi _1$ is
practically linear (see (\ref{div})), we may estimate 
\begin{equation}
\xi _c^{*}-\xi _1\simeq 1/g_{ef}^{\max }
\end{equation}
If we accept the cutoff (\ref{cut}), a small increase of $\xi ^{*}$ and $%
1/g^{*}$ will take place in an extremely narrow region near $g_c.$ This
increase cannot be practically observed.

In the 3D AFM case the statement about the finiteness of $\xi _c^{*}$ does
not, strictly speaking, hold since there exists a small region where $\Psi
(\xi )$ is positive and takes arbitrarily small positive values, so that the
decrease of $1/g_{ef}(\xi )$ can be slow. However, the behavior $\xi ^{*}(g)$
is in fact determined by the logarithmic singularity of the function $\Psi
(\xi )$ except for a very narrow region near $g_c.$ The numerical
calculations yield the estimation $g-g_c\sim 10^{-4}$ for the region where $%
\xi ^{*}$ starts to increase. Thus, from the practical point of view, we may
put $\xi _c^{*}$ to be finite. The corresponding value of $g_c^{\max }=g_1$
is determined by (\ref{ksic}) and $g_1^{*}$ is smaller than $g_{ef}^{\max }$.

At very small $g-g_c\sim 10^{-4}$ the value of $g^{*}$ starts to increase.
However, due to the minimum of the function $1/g_{ef}(\xi ),$ $%
g_c^{*}=1/\Delta $ remains finite at $g\rightarrow g_c,$ as well as in the
limit $N\rightarrow \infty .$ A similar situation takes place for
anisotropic ferromagnets.

In the case of an isotropic ferromagnet the influence of the singularity is
somewhat weaker since $\Psi (\xi )$ does not change its sign and takes small
positive values up to infinity, so that $\xi ^{*}$ starts to increase
appreciably at $g-g_c\sim 10^{-3}.$

The dependences $g_{ef}(\xi )$ in magnetic phases according to (\ref{gpsi})
at $g\rightarrow g_c\pm 0$ are shown in Figs.4-6.

Now we consider the results of the approach of Sect.8, which takes into
account the incoherent part of the spin spectral density. The integral of
motion of the system of equations (\ref{gll}), (\ref{wll}) has the same form
(\ref{int}), and we obtain 
\begin{equation}
\frac 1{Z(\xi )}=1+\frac 1a\ln \frac{\overline{\omega }}{\overline{\omega }%
_{ef}(\xi )}=1+\frac VN[g_{ef}(\xi )-g]  \label{1/Z}
\end{equation}
The corresponding equation for $g_{ef}$ reads 
\begin{eqnarray}
&&\ \ \frac{\partial (1/g_{ef})}{\partial \xi } 
\begin{array}{c}
=
\end{array}
-\Psi _{incoh}(\lambda +(aV/N)[g_{ef}-g]-\xi )  \label{g1psi} \\
&&\ \ \ \ \ \ \ \ \ \ -[\Psi _{coh}(\lambda +(aV/N)[g_{ef}-g]-\xi ) 
\nonumber \\
&&\ \ \ \ \ \ \ \ \ -\Psi _{incoh}(\lambda +(aV/N)[g_{ef}-g]-\xi )]/[1+\frac 
VN(g_{ef}-g)]  \nonumber
\end{eqnarray}
The role of the incoherent contribution becomes important only provided that 
$Z$ deviates appreciably from unity, i.e. $g_{ef}-g$ is large. This takes
place in a rather narrow region of $|g-g_c|.$ The estimation for $g_1$ in
the case of small $\delta $ reads now 
\begin{equation}
\frac{aVg_1^2/N}{1+Vg_1/N}\Psi _{coh}^{\max }=1
\end{equation}
so that, as follows from (\ref{gce}), $g_c$ increases. The dependences $%
g_{ef}(\xi )$ at small $|g-g_c|$ according to (\ref{g1psi}) are shown in
Fig.7. One can see that a crossover from the well-linear ``magnetic''
behavior to a PM-like ``quasi-linear'' behavior occurs with increasing $\xi
. $ The point of the crossover is estimated from $Z\Psi _{coh}^{\max }\simeq
1, $ i.e. 
\begin{equation}
(V/N)(\xi -\lambda )\simeq \Psi _{coh}^{\max }
\end{equation}
As demonstrate numerical calculations, the account of incoherent
contribution results in a smearing of the non-monotonous behavior of $%
g_{ef}(\xi )$ in the 3D AFM and anisotropic cases, and in some region of $%
|g-g_c|$ the minimum of $1/g_{ef}(\xi )$ vanishes completely. Therefore $%
g^{*}\rightarrow 0$ at $g\rightarrow g_c,$ unlike the situation for the
equation (\ref{gpsi}).

The influence of the incoherent contribution on $\xi ^{*}$ and $g^{*}$ is
considerably suppressed by the singularity of the function $\Psi _{coh}(\xi
).$ The region where this contribution starts to play a role is determined
by the quantity $\delta .$ In particular, for the 2D AFM case its influence
on $\xi ^{*}$ is practically absent since the divergence of $g_{ef}(\xi )$
occurs due to the singularity of $\Psi _{coh}(\xi ).$ Note that since $g^{*}$
is finite at $g\rightarrow g_c$, the coherent contribution survives up to $%
g_c$.

The comparison of the results of various approximations is presented in the
Table 1. One can see from this Table that for $N=2$ the relation of the $g_c$
values in the ordered phases an in the PM case is reversed in comparison
with the limit $N\rightarrow \infty .$ This fact is due to the influence of
the scaling function singularities. It should be noted that at larger $%
\delta \sim 1/5$ the value of $g_c$ in the ordered phases exceeds $%
\,g_c^{PM},$ as well as in the large-$N$ case. In the case (b) the value of $%
g_c$ is intermediate between $g_c^{(a)}\,$ and $g_c^{PM}$ and closer to $%
g_c^{(a)}$. With increasing $\alpha $ or $N$ the difference between $%
g_c^{(b)}\,$ and $g_c^{(a)}$ becomes still smaller.

Table 1. The critical values $g_c\,$ and $\xi _c^{*}$ for different magnetic
phases in the cases $N=\infty $ (see (\ref{g33}), (\ref{g44}), (\ref{t33}))
and $N=2$ in the approximation of Sect.7 (a) and with account of the
incoherent contribution (b). The parameter values are $\lambda =5,$ $\alpha
=1/2,\,\delta =1/100.$ For $N=2,$ the ``critical value'' of $\xi _c^{*}$ is
estimated from the plateau in the dependence $\xi ^{*}(g)$ (see the
discussion in the text).

\begin{tabular}{|l|l|l|l|l|l|}
\hline
&  & PM & FM & 3D AFM & 2D AFM \\ \hline
$N\to \infty $ & $g_c$ & 0.167 & 0.167 & 0.171 & 0.176 \\ \cline{2-6}
& $\xi _c^{*}$ & -- & -- & 5.35 & 5 \\ \hline
$N=2$ (a) & $g_c$ & 0.154 & 0.139 & 0.132 & 0.127 \\ \cline{2-6}
& $\xi _c^{*}$ & -- & 6.13 & 6.07 & 6.07 \\ \hline
$N=2$ (b) & $g_c$ & 0.154 & 0.141 & 0.136 & 0.131 \\ \cline{2-6}
& $\xi _c^{*}$ & -- & 6.23 & 6.17 & 6.16 \\ \hline
\end{tabular}

The dependences $1/g^{*}(g)$ and $\xi ^{*}(g)$ according to (\ref{g1psi})
are shown in Figs.8-11 (of course, these Figures do not show the
above-discussed increase of $1/g^{*}$ and $\xi ^{*}$ in the AFM case, which
takes place at very small $|g-g_c|$). The experimentally observable
quantities can be obtained from these data by using the formulas 
\begin{eqnarray}
S^{*} &=&\overline{S}_{ef}(C=0)=S\exp (-(V/N)[g^{*}-g]) \\
\overline{\omega }^{*} &=&\overline{\omega }_{ef}(C=0)=\overline{\omega }%
\exp (-(aV/N)[g^{*}-g])  \nonumber
\end{eqnarray}
($g<g_c$). For $g>g_c$ we have 
\begin{equation}
T_K^{*}=D\exp (-\xi ^{*})
\end{equation}
One can see from Figs.8-11 that, provided that $g$ is far from $g_c,$ we
have the one-impurity behavior $\xi ^{*}(g)\simeq 1/g,$ and the dependence $%
1/g^{*}(g)$ is given by (\ref{glin}), (\ref{gglin}), as well as in the large-%
$N$ limit.

\section{Critical behavior on the boundary of the strong-coupling regime:
breakdown of the Fermi-liquid picture}

To investigate the ``critical'' behavior of $\xi ^{*}$ at $g\rightarrow g_c$
we consider the function $1/g_{ef}(g,C)$ at small $|g-g_c|,|C|.$ First we
consider the results of the solution of the equation (\ref{gpsi}). With
approaching $g_c,$ the $\xi $-region, where the behavior (\ref{div}) takes
place, becomes very narrow and not too important for determining $\xi ^{*}.$
Unlike the exponential (in $\xi $) behavior in the large-$N$ limit, we have
from (\ref{linn}), 
\begin{equation}
g_{ef}(g=g_c,\xi \rightarrow \infty )\simeq -(N/Va)\xi  \label{Gc}
\end{equation}
However, the dependence $\xi ^{*}(g)$ turns out to be qualitatively the same
as for $N\rightarrow \infty $ (see (\ref{TKG})),

\begin{equation}
\xi ^{*}\simeq \gamma \ln (g-g_c)  \label{ksiln}
\end{equation}
Numerical calculations yield $\gamma =1/2$ for FM at $N=2,$ and $\gamma =1$
for FM with $N>2$ and PM. These values are the same as for $a\rightarrow 0$
(or according to the large-$N$ equation (\ref{intG}), provided that we take
for FM the function (\ref{g22g})). Thus one may put forward the hypothesis
that the critical exponents are universal, i.e. depend on the type of
magnetic ordering only, but not on $N,V$ and $\alpha $. In the AFM phases
numerical calculations yield the dependences (\ref{ksiln}) with $\gamma
\simeq 0.1$ ($d=3)$ and $\gamma \simeq 10^{-3}$ ($d=2).$

At the same time, the behavior of $g^{*}$ at $g\rightarrow g_c-0$ changes in
comparison with the large-$N$ limit. It turns out that for finite $N$ one
may establish a scaling relation of relevant variables at $g>g_c$ and $g<g_c$%
, as well as in the standard theory of critical phenomena. To find this
relation, we consider our problem in the region $|g-g_c|>\varepsilon $ where 
$\varepsilon \rightarrow 0$ determines a scale of approaching to the
critical point. When crossing the cut region, the argument of the function $%
\Psi $ should not shift considerably (Figs. 3-6). Indeed, this argument must
be close to zero; in the ordered phases it is fixed by the singularity
point, and for PM a considerable ``smearing'' takes place. Then we may
estimate 
\begin{equation}
\lambda +(aV/N)[g^{*}-g]\simeq \xi ^{*}
\end{equation}
so that 
\begin{equation}
g^{*}\simeq \gamma (N/Va)\ln (g_c-g)  \label{gln}
\end{equation}

As demonstrate numerical calculations (Fig.8), for the PM phase the increase
of $1/g^{*}$ at $g$ not too close to $g_c$ is almost linear in $g$, as well
as in the large $N$-limit, and the behavior (\ref{gln}) takes place only
starting from $g^{*}\sim 10$ which is, strictly speaking, beyond the
applicability of one-loop scaling. At the same time, for the FM case the
logarithmic dependence takes place in a considerable interval of $1/g^{*}$.
As discussed in previous Section, for 2D AFM the increase of $g^{*}$ and $%
\xi ^{*}$ at $g\rightarrow g_c$ can be hardly observed.

Of course, for 3D AFM $g_c^{*}=1/\Delta $ is in fact finite, and the
behavior (\ref{gln}) takes place at not too large $g^{*}.$ More exactly, we
can write down 
\begin{equation}
1/g^{*}\simeq 1/[\gamma (N/Va)\ln (g_c-g)]+\Delta
\end{equation}
However, practically the ``saturation'' region is extremely narrow and
cannot be achieved because of the smallness of $\gamma .$

When taking into account the incoherent contribution, a crossover to a
PM-like regime takes place at $g\rightarrow g_c$ in the dependences $%
g^{*}(g) $ and $\xi ^{*}(g)$, so that at very small $|g-g_c|$ we have the
behavior (\ref{ksiln}), (\ref{gln}) with $\gamma =1.$

Basing on the results (\ref{ksiln}), (\ref{gln}), it is natural to assume
that at $g\rightarrow g_c,$ $C\rightarrow -0$ one has the one-loop scaling
behavior 
\begin{equation}
g_{ef}(g,C)=-(N\gamma /aV)\ln (|C/\overline{\omega }|^{1/\gamma }+B(g_c-g)/g)
\label{sacal1}
\end{equation}
($B>0$ is a constant, the argument of the logarithm should be positive).
Then, according to (\ref{w+g}),(\ref{sw}) 
\begin{equation}
\overline{\omega }_{ef}(g,C)/\overline{\omega }\sim (|C/\overline{\omega }%
|^{1/\gamma }+B(g_c-g)/g)^\gamma  \label{wcc}
\end{equation}
\begin{equation}
\overline{S}_{ef}(g,C)/S\sim (|C/\overline{\omega }|^{1/\gamma
}+B(g_c-g)/g)^{\gamma /a}  \label{Gg}
\end{equation}
In particular, we have at $g\rightarrow g_c$ the power-law dependences 
\begin{eqnarray}
T_K^{*},\overline{\omega }^{*} &\sim &[\pm (g-g_c)]^\gamma ,  \label{wgam} \\
S^{*} &\sim &(g_c-g)^{\gamma /a}  \label{sgam}
\end{eqnarray}
Thus the ``critical exponents'' for the characteristic energy scales,
namely, $T_K^{*}$ at $g>g_c$ and $\overline{\omega }^{*}$ at $g<g_c$
coincide. Using (\ref{Gg}) we obtain at $g=g_c,$ $C\rightarrow 0$%
\begin{equation}
\frac{\overline{S}_{ef}(C)}S=\left( \frac{\overline{\omega }_{ef}(C)}{%
\overline{\omega }}\right) ^{1/a}\sim |C|^{\gamma /a}  \label{sscc}
\end{equation}

One of the most interesting consequences of our picture is a possibility of
a non-Fermi-liquid behavior on the boundary of the strong coupling region.
Indeed, during the renormalization process at $C\rightarrow 0$ the effective
spin-fluctuation frequency tends to zero, and the corresponding spectral
density is concentrated near $\omega =0.$ In this sense, the situation is
close to that in the one-impurity two-channel Kondo problem where a
collective mode with zero frequency occurs, which leads to a breakdown of
the Fermi-liquid picture due to electron scattering by this ``ultrasoft''
mode\cite{Col}. Unfortunately, our perturbation approach does not permit to
determine explicitly the temperature dependences of observables since the
coupling constant is not small in this regime. However, the calculations may
be performed within the large-$[l]$ $s-f$ model (see the next Section).

Of course, vanishing of $\overline{\omega }_{ef}(C)$ at $|C|=T_K^{*}$ and of 
$T_K^{*}$ and $\overline{\omega }^{*}$ at $g=g_c$ is the result of the
one-loop scaling, i.e. of using the lowest-order perturbation theory at
derivation of the renormalization group equations. In fact, one may expect
that in the strong-coupling region $\overline{\omega }_{ef}(C)\sim T_K$ \cite
{IKZ1,Von1}. One may assume that the correct scaling behavior, which may be
continued into the strong-coupling region, differs from the one-loop
behavior by the replacement 
\begin{equation}
B(g-g_c)/g\rightarrow (T_K^{*}/\overline{\omega })^{1/\gamma }
\end{equation}
A scaling law, which is more general than (\ref{wcc}), could be expected to
have the form 
\begin{equation}
\overline{\omega }_{ef}(C)/\overline{\omega }=(T_K^{*}/\overline{\omega }%
)\phi (C/T_K^{*})  \label{sacal2}
\end{equation}
Then one has 
\begin{equation}
\overline{S}_{ef}(C)/S=\psi (\overline{\omega }_{ef}(C)/\overline{\omega })
\end{equation}
A detailed investigation of magnetic properties (in particular, of the
formation of small moments, which are characteristic for heavy-fermion
systems) reduces to determining an explicit form of the functions $\phi $
and $\psi $. This problem cannot be solved within perturbative approaches.

\section{The non-Fermi-liquid behavior in the degenerate $s-f$ model}

As we have seen in Sect.9, in the large-$N$ limit the renormalization of
magnetic characteristics is weak in comparison with that of electron
spectrum. An opposite situation occurs in the case of large $[l]$ where the
number of electron branches is much larger than that of spin-wave modes, so
that the renormalization of spin dynamics plays the crucial role.

It is instructive to consider the large-$l$ limit in the $s-f$ model ($N=2$)
with 
\[
\lbrack l]\rightarrow \infty ,\,g\rightarrow 0,\,[l]g^2/2=\widetilde{g}^2=%
\text{const} 
\]
Then the effective $s-f$ interaction is unrenormalized, $\widetilde{g}_{ef}=%
\widetilde{g}=$ const, and the scaling equation takes the form 
\begin{equation}
\frac{\partial \chi }{\partial \xi }=a\widetilde{g}^2\Psi (\lambda +\chi
-\xi )  \label{linf}
\end{equation}
where 
\[
\chi (\xi )=\ln \frac{\overline{\omega }}{\overline{\omega }_{ef}(\xi )} 
\]
When taking into account the incoherent contribution we obtain instead of (%
\ref{linf}) 
\begin{eqnarray}
\frac{\partial \chi }{\partial \xi } &=&a\widetilde{g}^2[Z\Psi
_{coh}(\lambda +\chi -\xi )+(1-Z)\Psi _{incoh}(\lambda +\chi -\xi )]
\label{wzll} \\
Z &=&1/(1+\chi /a)  \nonumber
\end{eqnarray}
By introducing the function 
\begin{equation}
\nu =\lambda +\chi -\xi =\ln \frac{|C|}{\overline{\omega }_{ef}}
\end{equation}
the equation (\ref{linf}) can be readily integrated to obtain 
\begin{equation}
\int_\nu ^\lambda \frac{d\xi ^{\prime }}{1-a\widetilde{g}^2\Psi (\xi
^{\prime })}=\xi
\end{equation}
However, a simple qualitative analysis can be performed immediately for both
the equations (\ref{linf}) and (\ref{wzll}).

In the PM phase we have 
\begin{equation}
\chi (\xi )\simeq a\widetilde{g}^2\xi  \label{chi}
\end{equation}
up to the point 
\begin{equation}
\xi _1=\frac \lambda {1-a\widetilde{g}^2}
\end{equation}
Thus a power-law behavior occurs 
\begin{equation}
\overline{\omega }_{ef}(C)\simeq \overline{\omega }(|C|/D)^\beta ,\,\,\beta
=a\widetilde{g}^2
\end{equation}
For $\xi >\xi _1,$ 
\begin{equation}
\chi (\xi )\simeq \chi (\xi _1)=\lambda a\widetilde{g}^2(1-a\widetilde{g}%
^2)^{-1}
\end{equation}
is practically constant.

Note that unlike the case $l=0,$ which was discussed in the previous
Section, the bare coupling constant $\widetilde{g}$ and the exponent $\beta $
can be sufficiently large. At $\widetilde{g}\sim 1,$ $\overline{\omega }%
_{ef}(\xi )$ decreases rather rapidly during the renormalization process.
Thus a ``soft mode'' situation occurs, which may lead to a NFL behavior.

To investigate modification of the electron spectrum we may calculate the
second-order perturbation theory corrections, which are formally small in $%
1/[l]$. Replacing $\overline{\omega }\rightarrow \overline{\omega }_{ef}(C)$
in the usual second-order result for the electron self-energy (cf. Ref.\cite
{IKloc}) and introducing the effective electron density of states at the
Fermi level $N_{ef}(C)$ we obtain 
\begin{equation}
N_{ef}(C)\sim \frac{Z(C)}{\overline{\omega }_{ef}(C)}\sim \left( \frac D{|C|}%
\right) ^\beta \ln \frac D{|C|}  \label{Nef}
\end{equation}
To investigate qualitatively temperature dependences we may replace $%
|C|\rightarrow T.$ Thus one may expect an essentially NFL behavior of the
electronic specific heat, 
\begin{equation}
C_e(T)\sim TN_{ef}(|C|\rightarrow T)\sim T^{1-\beta }\ln (D/T),  \label{C}
\end{equation}
magnetic susceptibility, 
\begin{equation}
\chi _m(T)\sim S_{ef}^2(|C|\rightarrow T)/T\sim 1/T^{1-\beta /a},
\label{chi1}
\end{equation}
transport properties etc.

In magnetically ordered phases, the situation for $\xi >\xi _1$ changes
since the singularity of $\Psi _{coh}(\xi )$ plays an important role.
Provided that $a\widetilde{g}^2\Psi _{coh}^{\max }>1,$ the argument of the
function $\Psi _{coh}$ at $\xi >\xi _1$ becomes almost constant, and we
obtain 
\begin{equation}
\chi (\xi )\simeq \xi -\lambda ,\;\overline{\omega }_{ef}(C)\simeq |C|.
\label{ccc}
\end{equation}
The behavior (\ref{ccc}) is similar to the dependence (\ref{wcc}) for finite 
$l,$ and corresponds to $g=g_c.$ In the case of equation (\ref{linf}), this
behavior takes place up to $\xi =\infty .$ On the other hand, an account of
the incoherent contribution results in that the increase of $\chi $ stops at 
$a\widetilde{g}^2\Psi _{coh}^{\max }=1/Z=1+\chi /a,$ i.e. at 
\begin{equation}
\xi _2=\lambda +\chi _{\max }=\lambda +a(a\widetilde{g}^2\Psi _{coh}^{\max
}-1)
\end{equation}
Thus the value of $\xi _2$ is determined by the quantity $\delta $. The
dependence $\chi (\xi )$ for a 2D antiferromagnet is shown in Fig.12. In the
presence of the incoherent contribution the region, where the dependence (%
\ref{ccc}) holds, is rather narrow (especially for not too small $\delta ).$
However, a more exact consideration of spin dynamics (rather than using the
spin diffusion approximation) may change considerably the results.

In the regime (\ref{ccc}) we have the result (\ref{Nef}) with $\beta =1$ and 
$D\rightarrow \overline{\omega }$, so that in the AFM case ($a=1$) one
obtains 
\begin{equation}
\chi _m(T)=\text{const},C_e(T)\sim \ln (\overline{\omega }/T)
\end{equation}

For finite, but large $[l]$ the picture discussed fails below 
\begin{equation}
T_K=D\exp (-[l]^{1/2}/\sqrt{2}\widetilde{g})
\end{equation}
However, the NFL behavior takes place in a wide temperature region $T_K\ll
T\leq \overline{\omega }.$ At $T<T_K$ the renormalization of $g_{ef}$
becomes important and, as discussed in the end of the previous Section, a
more complicated scaling behavior may take place.

As discussed in Section 11, a NFL behavior takes place even for $l=0$ for $%
g\simeq g_c.$ The NFL region becomes broader with increasing $[l].$ The
dependences $g_{ef}(\xi )$ for $l=3$ are shown in Fig.13. One can see that
the linear dependence with a small coefficient takes place up to $\xi \simeq
5,$ then this is changed by the linear ``coherent'' behavior which is
further smeared by the incoherent contribution.

\section{Conclusions}

In conclusion, we resume main results of our consideration and their
relation to properties of the anomalous $f$-systems, and discuss some
unsolved problems.

Three regimes are possible at $I<0$ depending on the relation between the
one-impurity Kondo temperature $T_K$ and the bare spin-fluctuation frequency 
$\overline{\omega }$:

(i) the strong coupling regime with $I_{ef}(C\rightarrow 0)\rightarrow
\infty $ where all the conduction electrons are bound into singlet states
and spin dynamics is suppressed. This regime is expected to occur provided
that $\overline{\omega }\ll T_K$.

(ii) the regime of a ``Kondo'' magnet with an appreciable, but not total
compensation of magnetic moments, which corresponds to small $|g-g_c|$.

(iii) the regime of ``usual'' magnets with small logarithmic corrections to
the ground state moment and $\overline{\omega }_{ef}$. (Note that the same
situation takes place at $I>0$ \cite{IKJP92}.)

The formation of magnetic state takes place at 
\begin{equation}
T_K^c\equiv D\exp (-1/g_c)=A\overline{\omega }\left( \frac{\overline{\omega }%
}D\right) ^{1/a_c-1}  \label{tkc}
\end{equation}
($a_c\equiv \lambda g_c,$ $A$ is of order of unity). For $N\rightarrow
\infty $ we have $a_c=1+O(1/\lambda )$ and the strong coupling region
boundary is determined by the condition $T_K=A\overline{\omega }.$ For
finite $N$ we always have $a_c<1$ (see Table 1) and, according to (\ref{tkc}%
), $T_K^c\ll \overline{\omega }$. Numerical solution of the scaling
equations for the case $N=2$ demonstrates a considerable dependence of the
critical value $g_c$ on the type of magnetic ordering, space dimensionality
and the structure of the magnon spectrum (presence of the gap).

It should be stressed that the Doniach criterion $g_c\simeq 0.4$ \cite{Don},
which was obtained for a very special case in a simplified one-dimensional
model, cannot in fact be used for real systems since $g_c$ turn out to be
sensitive to parameters of exchange interactions, type of magnetic ordering,
space dimensionality, degeneracy factors, magnetic anisotropy etc. It is
worthwhile to note in this connection that it is a common practice to treat
the interplay of spin dynamics and the Kondo effect within the two-impurity
problem (see, e.g., Refs.\cite{Abr}). At the same time, we have demonstrated
that the most important features of the scaling behavior are connected with
peculiarities of the spin spectral density and are not described by this
model (where the spectral density is a delta-like peak corresponding to
singlet-triplet transitions).

We have used in our calculations the simplest ``Debye'' approximation for
the magnon spectrum, i.e. the long-wave dispersion law in the whole
Brillouin zone. Competing interspin interactions owing to the oscillating
behavior of the RKKY exchange, which may lead to frustrations, might be also
important for explaining magnetic structures in ``usual'' Kondo lattices.

The effective Kondo temperature $T_K^{*}$ determines a characteristic energy
scale of the ``heavy-fermion'' behavior at low temperatures. This may differ
considerably from the one-impurity value $T_K$, so that in the PM phase $%
T_K^{*}<T_K,$ and in the magnetically ordered phases, except for very small $%
g-g_c,$ $T_K^{*}>T_K$ (see Figs.9-11). In the AFM case $T_K^{*}$ depends
weakly on $g$ and therefore on $T_K,$ and is determined by $\overline{\omega 
}$ in a wide interval of $g.$

Although our consideration was performed for $T=0,$ one may expect by the
analogy with the one-impurity problem that the dependences $\overline{S}(T)$
(in the PM phase, $\overline{S}(T)$ is the local moment determined from the
magnetic susceptibility) and $\overline{\omega }(T)$ may be qualitatively
obtained by the simple replacement $|C|\rightarrow T.$

For $N=2$ and small enough $\delta $ (spin excitation damping) the Kondo
screening in the AFM and FM phases is stronger than in the PM phase. This
results in an increase of the Rhodes-Wohlfarth ratio (ratio of the effective
moment, as determined from the Curie constant, to the saturation moment).
For example, if the bare coupling constant is close to its critical value in
the ordered phase and correspondingly lower than in the PM phase, the ground
state moment may be small in comparison with the high-temperature one (a
behavior, typical for most Kondo magnets, as well as weak itinerant electron 
$d$-magnets). The same situation takes place for the characteristic
spin-fluctuation frequency $\overline{\omega }_{ef}.$ It should be noted
that an increase of spin-fluctuation energies with temperature is indeed
observed in a number of anomalous $f$-systems, e.g., U$_2$Zn$_{17}$ \cite
{UZn}.

Presence of the factor $1/N$ in the AFM2 case or large value of $\delta $
(magnon damping) may result in that the suppression of the effective moment
and spin-dynamics frequency turns out to be weaker than in the PM phase. It
looks like the decrease of moment at magnetic disordering with increasing
temperature. Such a decrease is typical for strong itinerant magnets where
it is due to the change of electron spectrum at disordering (see Refs.\cite
{VKT,IKUFN}). In the case under consideration this phenomenon has a quite
different (essentially many-particle) nature.

Near the boundary of the strong-coupling region ($g\rightarrow g_c$) the
relevant variables demonstrate a non-trivial scaling behavior as functions
of $|g-g_c|$ and $\xi .$ In particular, 
\begin{equation}
T_K^{*}(g\rightarrow g_c+0),\overline{\omega }^{*}(g\rightarrow g_c-0)\sim
|g-g_c|^\gamma ,
\end{equation}
the exponent $\gamma $ depending on the type of magnetic ordering. These
results may be of interest for the general theory of metallic magnetism. The
description of the state with small magnetic moments ($g\rightarrow g_c$)
turn out to differ considerably from that in the theory of weak itinerant
magnetism \cite{Mor}. It is interesting that the ``critical exponents'' in
the dependences of the moment on the coupling constant (\ref{sgam}) and $C$ (%
\ref{sscc}) turn out to be non-integer. The corresponding dependences $%
\overline{S}(T)=\overline{S}_{ef}(|C|\rightarrow T)$ describe an analogue of
the ``temperature-induced magnetism'' \cite{Mor}.

As mentioned in the Introduction, high sensitivity of the magnetic state in
heavy-fermion systems to external factors is explained by that in the case
(ii) the magnetic moment changes strongly at small variations of the bare
coupling constant. According to our consideration, the regime of small
magnetic moments occurs in a very narrow region of bare parameters only.
Provided that $g$ is not too close to $g_c,$ a characteristic interval of
the change of the quantity $g^{*}$ (which determines the renormalized values
of magnetic moment and spin-fluctuation frequency) by unity is estimated as $%
\delta g\sim g^2\ll g$. In the immediate vicinity of $g_c$ (where the
behavior $g^{*}\sim -\ln |g-g_c|$ takes place) this interval becomes still
more narrow: $\delta g\sim |g-g_c|.$ A more consistent treatment of this
regime with account of possible renormalization of the scale $|g-g_c|$
itself requires using more complicated (e.g., numerical) scaling approaches.

``Softening'' of the spin excitation spectrum in the critical region may
result in a non-Fermi-liquid behavior. Although the NFL state itself cannot
be described within the framework of our perturbative approach, the
conclusion about the NFL behavior near the boundary of magnetic ($g<g_c$)
and non-magnetic ($g>g_c$) phases seems to be important. This conclusion is
confirmed by the fact that a violation of the Fermi-liquid picture in
anomalous $f$-system is really observed near the onset of magnetic ordering 
\cite{Maple}. It is difficult to explain this fact within the frequently
used one-impurity two-channel Kondo model\cite{Tsv}. At the same time, our
scenario of the NFL state formation takes into account essentially $%
many-center$ nature of the system. The width of the region where the NFL
behavior occurs increases with increasing the degeneracy factor $[l]$ and
decreases with increasing $N.$ In the formal limit $[l]\rightarrow \infty $
the perturbation theory in $g$ remains applicable, so that explicit
expressions for thermodynamic and magnetic properties can be obtained.

On the whole, the physical behavior, which occurs as a result of an
interplay between the Kondo effect and intersite exchange interactions,
turns out to be very rich and differs for various model versions.

The work was supported in part by Grant 96-02-16000 from the Russian Basic
Research Foundation.

{\sc Figure captions}

Fig.1. The dependence $1/g_{ef}$ on $\xi =\ln |D/C|$ in the large-$N$ limit
with $\lambda =\ln (D/\overline{\omega })=5,\,g=0.15$ for a paramagnet
(dashed line) and different magnetic phases (solid lines): (1) ferromagnet
(2) 3D antiferromagnet (3) 2D antiferromagnet (4) magnet with a strong
anisotropy $\omega _0\rightarrow \infty $.

Fig.2. The dependence $1/g_{ef}(\xi )$ according to the equation (\ref{gpsi}%
) at $N=2$ in the anisotropic case ($\omega _0=0.2$) for different magnetic
phases: (1) ferromagnet (2) 3D antiferromagnet (3) 2D antiferromagnet, and
for the isotropic ferromagnet (dotted line) and 3D antiferromagnet (dashed
line). The bare coupling parameter is $g=0.11,$ other parameters are the
same as in Sect.10.

Fig.3. The scaling trajectories $g_{ef}(\xi )$ in a paramagnet according to (%
\ref{gpsi}) with $N=2$ for $g=0.153917>g_c$ (upper line) and $g=0.153916<g_c$
(lower line).

Fig.4. The scaling trajectories $g_{ef}(\xi )$ in a ferromagnet according to
(\ref{gpsi}) with $N=2$ for $g=0.13868>g_c$ (upper line) and $g=0.13867<g_c$
(lower line).

Fig.5. The scaling trajectories $g_{ef}(\xi )$ in a 3D antiferromagnet
according to (\ref{gpsi}) with $N=2$ for $g=0.1320382>g_c$ and $%
g=0.1320381<g_c$ .

Fig.6. The scaling trajectories $g_{ef}(\xi )$ in a 2D antiferromagnet
according to (\ref{gpsi}) with $N=2$ for $g=0.1266714>g_c$ and $%
g=0.1266713<g_c$.

Fig.7. The scaling trajectories $g_{ef}(\xi )$ in a 3D antiferromagnet
according to (\ref{g1psi}) with $N=2$ for $g=0.136305>g_c$ (upper line) and $%
g=0.136305<g_c$ (lower line).

Fig.8. The dependences $1/g^{*}(g)$ for $g<g_c$ and $\xi ^{*}(g)-\lambda $
for $g>g_c$ in a paramagnet ($\lambda =5)$. The dashed line is the curve $%
1/g-\lambda $.

Fig.9. The dependences $1/g^{*}(g)$ for $g<g_c$ and $\xi ^{*}(g)-\lambda $
for $g>g_c$ in a ferromagnet according to (\ref{g1psi}) The parameters are $%
\lambda =5,$ $\delta =1/100$.

Fig.10. The dependences $1/g^{*}(g)$ for $g<g_c$ and $\xi ^{*}(g)-\lambda $
for $g>g_c$ in a 3D antiferromagnet according to (\ref{g1psi}).

Fig.11. The dependences $1/g^{*}(g)$ for $g<g_c$ and $\xi ^{*}(g)-\lambda $
for $g>g_c$ in a 2D antiferromagnet according to (\ref{g1psi}).

Fig.12. The dependence of $\chi =\ln (\overline{\omega }/\overline{\omega }%
_{ef})$ vs. $\xi =\ln |D/C|$ for a 2D antiferromagnet at $[l]=\infty $ with $%
\delta =10^{-3},$ $\lambda =5$, $\widetilde{g}=0.6)$ according to (\ref{linf}%
) (dashed line) and with account of the incoherent contribution (solid line).

Fig.13. The scaling trajectories $g_{ef}(\xi )$ in a 3D antiferromagnet
according to (\ref{g1psi}) with $N=2,$ $l=3,$ $\delta =1/100$ for $%
g=0.10556>g_c$ (upper line) and $g=0.10555<g_c$ (lower line).


\begin{references}
\bibitem[*]{E}  E-mail: Mikhail.Katsnelson@usu.ru

\bibitem{Stewart}  G.R.Stewart, Rev.Mod.Phys.{\bf 56, }755 (1987).

\bibitem{Brandt}  N.B.Brandt and V.V.Moshchalkov, Adv.Phys.{\bf 33, }373
(1984); V.V.Moshchalkov and N.B.Brandt, Usp.Fiz.Nauk, {\bf 149}, 585 (1986).

\bibitem{Adr}  D.T.Adroya and S.K.Malik, J.Magn.Magn.Mat.{\bf 100}, 126
(1991) (see also references therein).

\bibitem{Fulde}  P.Fulde, J.Keller, and G.Zwicknagl, Solid State Physics,
vol.41, ed.F.Seitz et al., New York, Academic Press, 1988, p.1.

\bibitem{Zw}  G.Zwicknagl, Adv.Phys.{\bf 41}, 203 (1992).

\bibitem{Lee}  P.A.Lee, T.M.Rice, J.W.Serene, L.J.Sham, and J.W.Wilkins,
Comm.Cond.Mat.Phys.{\bf 12}, 99 (1986).

\bibitem{Lac}  C.Lacroix, J.Magn.Magn.Mat.{\bf 100, }90 (1991).

\bibitem{Don}  S.Doniach, Physica B{\bf 91, }231 (1977).

\bibitem{IKFTT}  V.Yu.Irkhin and M.I.Katsnelson, Fiz.Tverd.Tela {\bf 30},
2273 (1988); V.Yu.Irkhin and M.I.Katsnelson, Z.Phys.B{\bf 75}, 67 (1989).

\bibitem{IKZ1}  V.Yu.Irkhin and M.I.Katsnelson, Z.Phys.B{\bf 82}, 77 (1991).

\bibitem{Von1}  V.Yu.Irkhin and M.I.Katsnelson, Fiz.Met.Metalloved.No1, 16
(1991) [Phys.Met.Metallography {\bf 71}, 13 (1991)]; S.V.Vonsovsky,
V.Yu.Irkhin, and M.I.Katsnelson, Physica B{\bf 163, }321 (1990).

\bibitem{Von2}  S.V.Vonsovsky, V.Yu.Irkhin, and M.I.Katsnelson, Physica B%
{\bf 171}, 135 (1991)

\bibitem{IKJP92}  V.Yu.Irkhin and M.I.Katsnelson, J.Phys.:Cond.Mat.{\bf 4},
9661 (1992).

\bibitem{Maple}  M.B.Maple et al, J.Low Temp.Phys.{\bf 95}, 225 (1994); {\bf %
99}, 223 (1995).

\bibitem{CeNi}  K.Umeo, H.Kadomatsu, and T.Takabatake, J.Phys.:Cond.Mat.{\bf %
8}, 9743 (1996).

\bibitem{Steg}  F.Steglich et al, J.Phys.:Cond.Mat.{\bf 8}, 9909 (1996).

\bibitem{Tsv}  B.Andraka and A.M.Tsvelik, Phys.Rev.Lett.{\bf 67}, 2886
(1991); A.M.Tsvelick and M.Rivier, Phys.Rev.B{\bf 48}, 9887 (1993).

\bibitem{Col}  P.Coleman, L.B.Ioffe, and A.M.Tsvelik, Phys.Rev.B{\bf 52},
6611 (1995).

\bibitem{PL}  V.Yu.Irkhin and M.I.Katsnelson, Pis'ma ZhETF {\bf 49}, 550
(1989); Phys.Lett.A{\bf 150, }47 (1990).

\bibitem{Noz}  P.Nozieres and A.Blandin, J.Phys.(Paris) {\bf 41}, 193 (1980)

\bibitem{Wieg}  A.M.Tsvelick and P.B.Wiegmann, Adv.Phys.{\bf 32}, 745 (1983).

\bibitem{CS}  B.Coqblin and J.R.Schrieffer, Phys.Rev.{\bf 185}, 847 (1969).

\bibitem{Schr}  J.R.Schrieffer, J.Appl.Phys. {\bf 38}, 1143 (1967).

\bibitem{Hirst}  L.L.Hirst, Z.Phys.{\bf 244}, 230 (1971); Int.J.Magn.{\bf 2}%
, 213 (1972); Adv.Phys.{\bf 21}, 295 (1972).

\bibitem{II}  V.Yu.Irkhin and Yu.P.Irkhin, Zh.Eksp.Theor.Phys.{\bf 107}, 616
(1995).

\bibitem{Orb}  H.Brooks, Phys.Rev.{\bf 58, }909 (1940).

\bibitem{VKT}  S.V.Vonsovsky, A.V.Trefilov, and M.I.Katsnelson,
Phys.Met.Metallography {\bf 76}, 247, 343 (1993).

\bibitem{New}  D.M.Newns and N.Read, Adv.Phys.{\bf 36}, 799 (1987).

\bibitem{Bick}  N.E.Bickers, Rev.Mod.Phys.{\bf 59}, 845 (1987).

\bibitem{FLow}  P.Fulde and M.Loewenhaupt, Adv.Phys.{\bf 34}, 589 (1986).

\bibitem{Cox}  T.-S.Kim and D.L.Cox, Phys.Rev.B{\bf 54}, 6494 (1996).

\bibitem{IK96}  V.Yu.Irkhin and M.I.Katsnelson, Phys.Rev.B{\bf 53}, 14008
(1996).

\bibitem{Aus}  M.I.Auslender and V.Yu.Irkhin, Z.Phys.B{\bf 56}, 301 (1984).

\bibitem{And}  P.W.Anderson, J.Phys.C{\bf 3}, 2346 (1970).

\bibitem{Mor}  T.Moriya, Spin Fluctuations in Itinerant Electron Magnetism,
Springer, Berlin, 1985.

\bibitem{IKloc}  V.Yu.Irkhin and M.I.Katsnelson, Z.Phys.B{\bf 70}, 371
(1988).

\bibitem{Abr}  E.Abrahams, J.Magn.Magn.Mat.{\bf 63-64, }235 (1987);
T.Yanagisawa, J.Phys.Soc.Jpn, {\bf 60}, 3449 (1991).

\bibitem{UZn}  G.Broholm, J.K.Kjems, and G.Aeppli, Phys.Rev.Lett.{\bf 58, }%
917 (1987).

\bibitem{IKUFN}  V.Yu.Irkhin and M.I.Katsnelson, Usp.Fiz.Nauk {\bf 164,} 705
(1994).
\end{references}
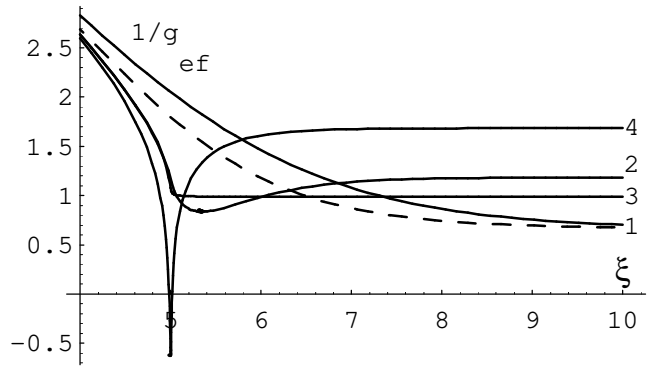
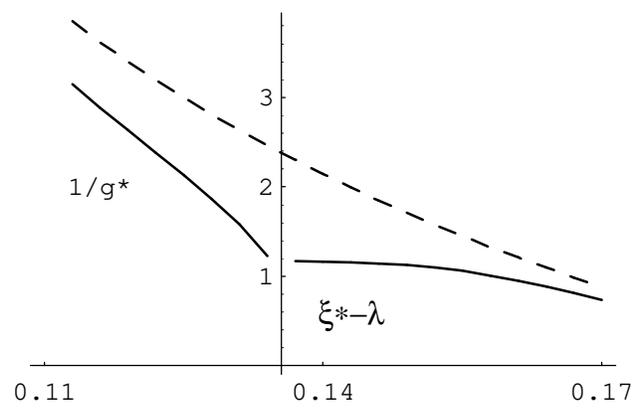
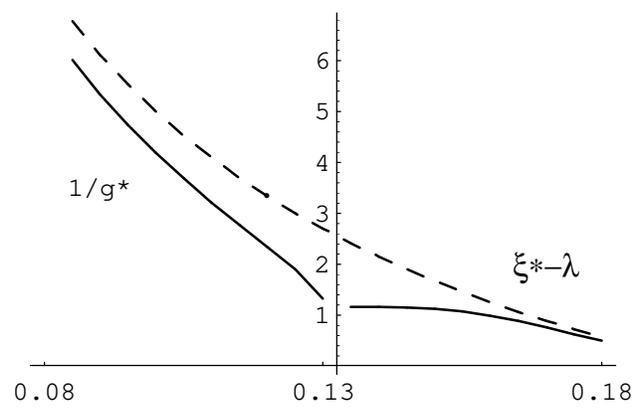
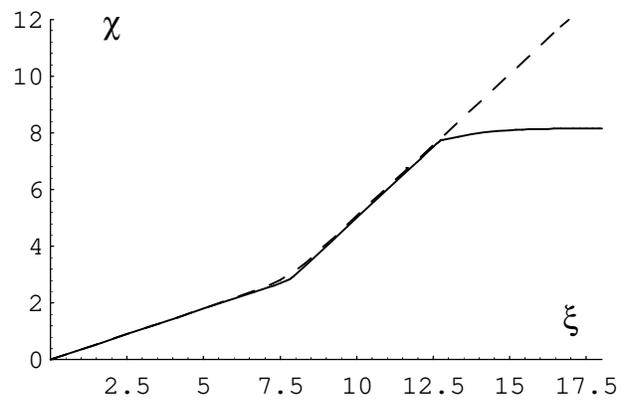
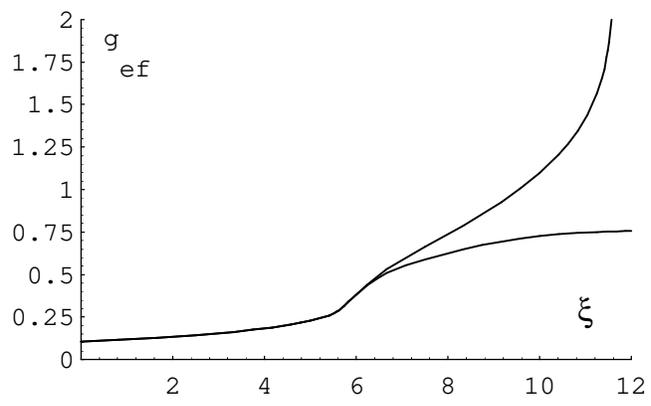
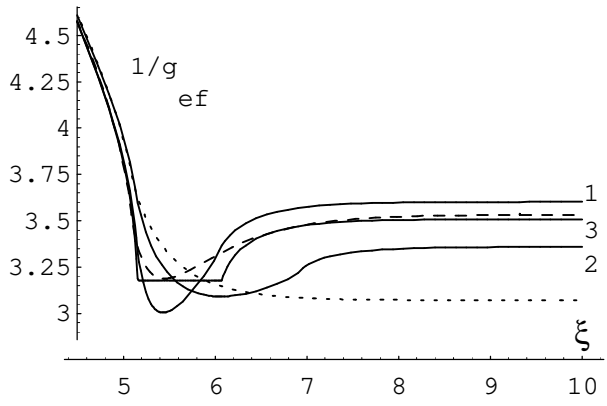
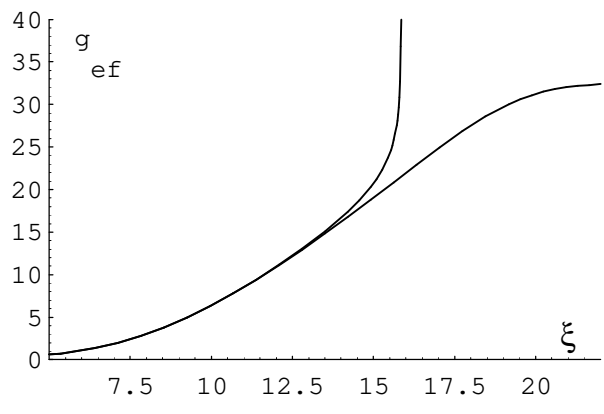
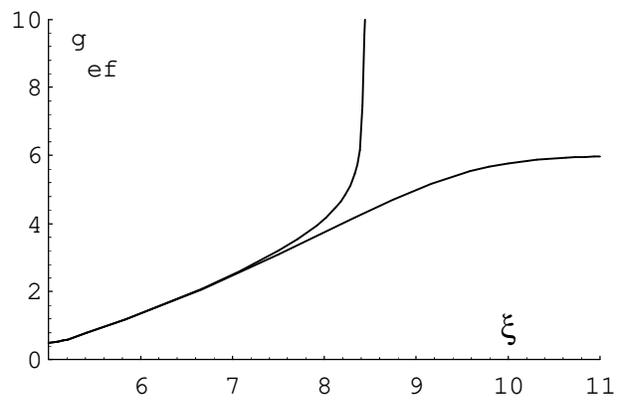
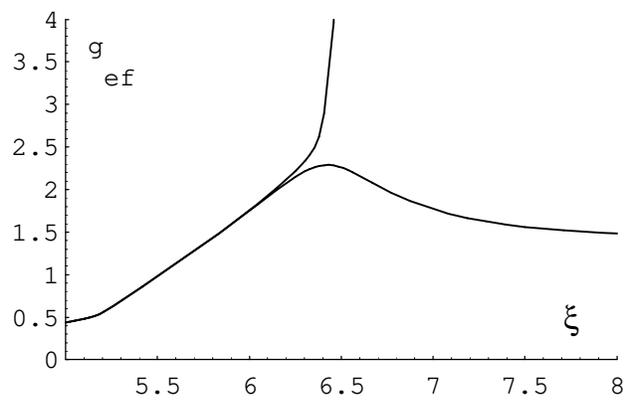
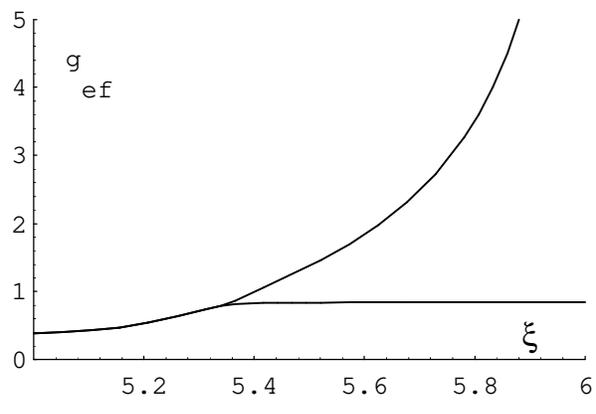
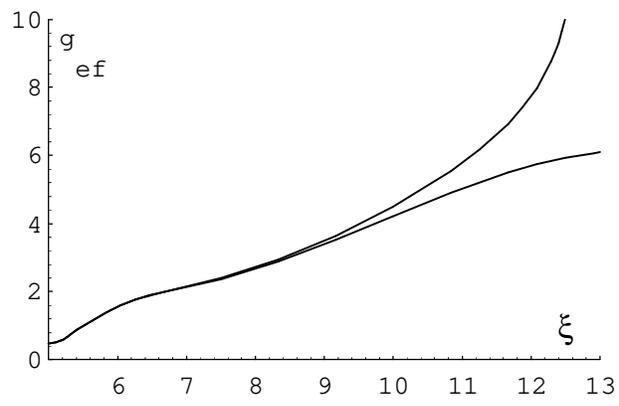
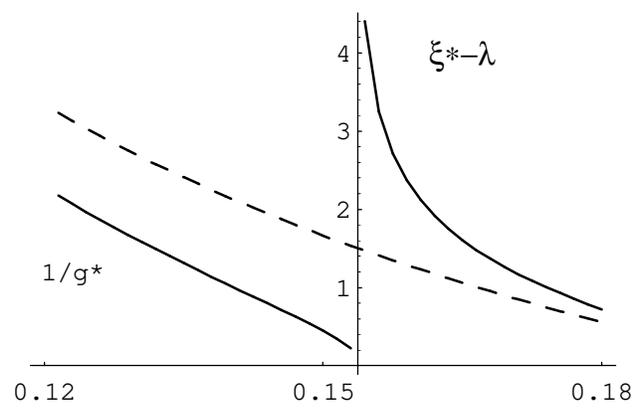
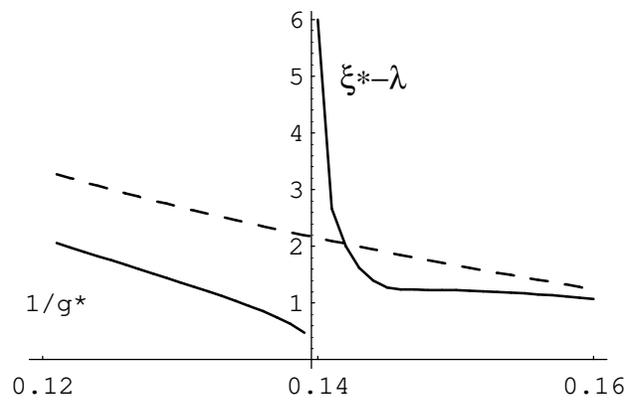
\end{document}